\documentclass[a4paper,landscape,10pt]{article}

\usepackage{graphicx}
\usepackage{verbatim}
\usepackage{amssymb}
\usepackage{amsmath}
\usepackage{color}
\usepackage[parfill]{parskip}
\usepackage[margin=2.5cm]{geometry}
\usepackage{url}
\usepackage{hyperref}
\usepackage{graphicx}
\usepackage{algorithm,algpseudocode}
\usepackage{placeins}
\usepackage{setspace}
\usepackage{fancyhdr}
\usepackage{graphicx,graphics}
\usepackage{amsmath, amsthm, amssymb}
\usepackage{anysize}\usepackage{subfig}
\usepackage{multirow}\usepackage{array}
\usepackage{xcolor}\usepackage{url}

\begin{document}

\title{Real-time Ionospheric Imaging of $S_4$ Scintillation from Limited Data with  Parallel Kalman Filters and Smoothness}

\author{Alexandra~Koulouri
\thanks{A. Koulouri  was with the Faculty of Information Technology and Communication Sciences,
Tampere University, P.O. Box 692, 33101 Tampere, FI,  e-mail: alexandra.koulouri@tuni.fi.}% <-this % stops a space
}

\maketitle

\begin{abstract}
In this paper, we propose a Bayesian framework to create two dimensional ionospheric images  of high spatio-temporal resolution to monitor ionospheric irregularities as measured by the $S_4$ index. % scintillation maps. % In this paper, we propose a Bayesian framework to cre-ate high spatio-temporal resolution two dimensional scintillationmaps. 
Here, we recast the standard Bayesian recursive filtering for a linear Gaussian state-space model, also referred to as the Kalman filter, first by  augmenting the (pierce point) observation model with connectivity information stemming from the insight and assumptions/standard modeling about the spatial distribution of the scintillation activity on the ionospheric shell at 350 km altitude. Thus, we achieve to handle the limited  spatio-temporal observations. Then, by introducing a set of Kalman filters running in parallel, we mitigate the uncertainty related to 
a tuning parameter
 of the proposed augmented model. The output images  are a weighted average of the state
estimates of the individual filters.
We demonstrate our approach by rendering two dimensional real-time ionospheric images of $S_4$ amplitude scintillation at 350 km over South America with  temporal resolution of one minute. Furthermore, we employ extra $S_4$ data that was not used in producing these ionospheric images, to check and verify the ability of our images to predict this extra data in particular ionospheric pierce points. Our results show that in areas with a network of ground receivers with a relatively good coverage (e.g. within  a couple of kilometers distance) the produced images can provide reliable real-time results. Our proposed
algorithmic framework can be
readily used to visualize real-time ionospheric images taking
as inputs the available scintillation data provided from freely available web-servers.
%Our proposed approach can readily be applied in imaging other ionospheric scintillation indices such as phase scintillation. 
\end{abstract}

%\begin{keywords}
%Real-time ionospheric imaging, scintillation, $S_4$ index, GNSS system, Bayesian filtering, discrete Kalman, smoothness effect, ensemble of filters, tuning parameter
%\end{keywords}


%MAIN ARTICLE

\section{Introduction}

The discrete time Kalman filter \cite{Kalman1960,Kalman1961} is a
widely applied recursive Bayesian approach for multivariate normal distributions that allows to combine instantaneous
measurements, spatial and temporal prior information to obtain
estimates for a dynamically evolving system. Kalman filter, even
though it was developed originally for tracking targets such as
aircrafts \cite{Alsadik2019}, today  has found applications in many
research areas in engineering and signal processing
\cite{Auger2013} such as speech
recognition\cite{Gannota}, tomography \cite{Vauhkonen1998,Daun2011},
hydrology \cite{Clark2008}  and econometrics \cite{Wilcox2017} to
name but a few.

The Earth's ionospheric weather, affected by the solar activity, is
a purely dynamic system whose turbulence has a significant impact on
 navigation, positioning and  satellite communication systems
that are an integral part of many human activities
\cite{Shi2019,Awange2012}. 
Either Empirical \cite{Bilitza1988} or physical-based models (e.g.
\cite{Angling2011}), combined with data assimilation techniques
\cite{Elvidge2019}, have been developed to track and visualize the
global ionospheric climate and weather.
In particular,  Kalman filtering
 has been proposed for example in
\cite{Hsu2018,Codrescu2018,Sutton2018,Elvidge2019,He2020} to calibrate
coupled ionospheric-thermospheric physical models in order to
monitor and forecast the ionospheric and plasma-spheric electron
density distributions and in tomographic recovery of ionospheric total
electron content (TEC).

Even though, electron density or total electron content (TEC) tomographic images can provide useful
information about the ionospheric conditions, numerous ionospheric
phenomena remain untracked due to the complexity and variability of
the ionospheric environment \cite{Sun2019}. For example, at lower
latitudes, satellite communications using frequencies up to a few GHz
can experience significant signal amplitude and phase fluctuations
due to small-scale time-varying ionospheric plasma density
irregularities, a phenomenon known as
 scintillation \cite{Crane1977,Aarons1982,Aquino2005}.
The amplitude scintillation (which is manifested as random rapid
fluctuations in the signal strength) can cause periods of reduced
signal power at the receiver's antenna that can result in a power
drop below the receiver-tracking threshold (loss of lock incidents)
\cite{Jiao2015}. The continuous monitoring of scintillation activity
especially in the equatorial area \cite{Paul2010,Jiao2016,Jiao2018a}
is substantial to mitigate
potential risks 
primarily for safety critical activities that rely on satellite
communications. At the moment, this is supported by a distributed network of
scintillation monitors (receivers) over South America for instance.
Unfortunately, the existing network can provide scattered and sparse
information about the activity over the continent. Moreover, climatology models or other numerical approaches,
e.g. \cite{Spogli2009,Li2010,Jacobsen2014,Alfonsi2018,Koulouri2020}, struggle to produce
 real time pictures that reflect the small-scale ionospheric plasma density fluctuations  due to their
design or inherent limitations. For example, models may break down in the presence of strong scintillation activity \cite{Priyadarshi2015} whereas numerical approaches often rely on collecting data over 
%and  often, it is standard to
%constructing static maps at the 350 km ionospheric shell using projected %scintillation data collected over 
 long periods of time and then illustrate
 an average or gross scintillation activity (e.g. averaged $S_4$ values) that cannot capture the scintillation spikes,
 spatial extent
 and  dynamic evolution of the phenomenon.
Here, we aim to address these limitations and  create instantaneous
ionospheric images that reflect the dynamically changing scintillation
activity with the help of an enriched Kalman filter framework. To the best
of our knowledge, this is the first time that 
$S_4$ amplitude/intensity ionospheric scintillation images of high spatio-temporal resolution  are produced.

\subsection{Our contributions}
  %\item 
In this work, we formulate a state-space problem \cite{Murphy2012,Fruehwirth-Schnatter2006} that describes
macroscopically the evolution of the ionospheric scintillation activity (as measured by the $S_4$ index). Particularly, at each time step a standard transition model is used to express the system's dynamics which is then corrected using a developed observation
model which employs the available scintillation measurements.
 Our
main contributions are as follows.
\begin{itemize}
\item We develop an observation model for the filter with the help of the
finite element method \cite{Hughes2000}. Here, the measured
values at pierced points are connected with a set of
distributed scintillation values at fixed
  locations (called nodes)  through linear basis functions.
  % on the ionospheric shell at 350km
  Hence, we produce the images on a mesh described by nodes and elements (instead of a uniform grid). %to estimate the map's
  %values  where
  The   elements' size and shape depend on the traces of the available
 observations on the ionospheric shell at 350 km altitude where the ionospheric images are rendered \cite{Davies1990}.
The use of linear
  basis functions compared to piece-wise constant functions (i.e. fixed values) as in a regular grid allows smooth
  transitions between neighboring nodes on the image and the capture of smaller scintillation
  fluctuations. %Furthermore, a mesh allows to enclose only the relevant ionospheric area. % with available data.
  \item We augment the observation model by employing a connectivity (smoothness) prior \cite{2005Kaipio} (i.e. a set of extra equations) which is regularized
by a tuning parameter that allow us to handle
the limited available data  at each time instant.
\item Since, a Kalman filter requires prior knowledge about the model
parameters for optimal performance \cite{Price1968,Fitzgerald1971},
we handle the tuning parameter choice by employing a bank  of Kalman filters (here called ensemble)
 which run in parallel \cite{Magill1965,Blom1988,Li1993,Chaer1997}.
In particular, each member of the ensemble is modeled with a different
realization for the tuning parameter (for the connectivity
prior) selected from a fixed set. The output (a.k.a. the $S_4$ image) at
each step is a weighted average of the individual filter state
estimates at that time step. %Our weighted average (which is an approximate inference at each time step) 
Our proposed scheme is based on the application of a generalized technique called assumed density filtering \cite{Maybeck1979}.
The weights are scalar quantities which
are estimated (on-line) based on the performance of the individual
filters determined using control $S_4$ measurements  at
each step. Hence,
the filters with the highest weights represent the
ones with the ``optimal'' parameter values. 
Compared to \cite{Sims1969} where a weight is computed according to
the performance over the entire sequence of available control
measurements, here, we use only the fraction of the control
data that corresponds at that particular time instant. % i.e. we employ only the most recent control
 %measurements. 
With this scheme, we allow the weights to
adjust quickly based on the most recent measurements. Moreover, we avoid
underflowing problems (i.e. cases where weights tend to zero due to recursive
multiplication of exponents)\cite{Sims1969}.
\item We develop an algorithm that uses the proposed enriched Kalman filter framework and we produce real time images of $S_4$ value over South America. 
\end{itemize}
We remark that the standard Kalman filter has been previously applied to reduce or
mitigate scintillation effects in Global Navigation Satellite
Systems (GNSS) signal tracking or precise point positioning see e.g.
\cite{Barreau2012,VilaValls2018,Veettil2020} and references
therein. However, the developed approach, as well as the problems in question here are totally different.
In particular, our work's novelties, which are summarized to the
development of an augmented state-space model that describes the
statio-temporal evolution of the scintillation activity
and then the design of an ensemble of Kalman filters (in order to
optimise the level of smoothness of the images), % reduce modelling uncertainties) %it has not be used to produce dynamic maps that
%visualize the scintillation activity (e.g. over the South America).
%The framework, proposed in this works, allows
allow us to create dynamically evolving ionospheric images e.g. images of
$S_4$ values over South America as we show in the results sections \ref{section:results} and \ref{section:Valid}. These can readily be used to observe in real-time the ionospheric
changes due to scintillation,  can be further employed as a
prior information in the estimation of dilution of precision metrics (e.g. in conjunction with the approach presented in
\cite{Koulouri2020}), or enrich climatology databases. We note that
the developed algorithmic framework and the accompanied software can
be readily used to create on-line ionospheric images taking
  as inputs available scintillation data provided for example
  from a server such as   CIGALA/CALIBRA network-
UNESP web server, (ISMR Query Tool \cite{Vani2017}).
In this work, we demonstrate the proposed approach by showing a
sequence of  $S_4$ images and providing a video (\href{https://www.youtube.com/watch?v=K8BJFWTLzmw}{link}). Finally, we validate the
accuracy of the produced images by comparing our predictions with held-out
measurements that were not included in the image estimation. We note that our approach can be  readily applied in imaging other ionospheric scintillation values such as phase scintillation for instance. 
%%%%%%%%%%%%%%%%%%%%%%%%%%%%%%%%%%%%%%%%%%

\section{Theory}\label{sec:Theory}

\subsection{Observation model for the ionospheric scintillation}\label{sec:ModellingIonScMaps}
%The set of the linear equations
%(eq.~\ref{eq:LinearEquationsIonosphere}) can be expressed in
%vector-matrix form as
Let us start with the model that connects the monitored
scintillation data with the ionospheric scintillation activity
\begin{equation}\label{eq:ObservationModel}
y_t = A_t s_t+\varepsilon_t,
\end{equation}
where $y_t\in \mathbb{R}^{M_t}$ are the measured data,
$s_t\in\mathbb{R}^N$ is the vector that encloses the ionospheric
scintillation distribution to be estimated, $A_t\in
\mathbb{R}^{M_t\times N}$ is the design matrix and $\varepsilon_t\in
\mathbb{R}^{M_t}$ is the additive noise which here
 is modelled as Gaussian with zero mean and covariance
$\Gamma_{\varepsilon_t}$ i.e.
$\varepsilon_t\sim\mathcal{N}(0,\Gamma_{\varepsilon_t})$. In this
problem, the size $M_t$ of the instantaneous $y_t$ is far less than
the unknown discrete distribution of $s_t$ that is required to
produce an instantaneous ionospheric image (i.e. $M_t\ll N $),
therefore prior information is needed for the inversion. In the
following section, we employ the discrete Kalman filter to
combine instantaneous measurements with priors
to estimate  high resolution ionospheric scintillation images at $t$
time steps.

\subsection{Kalman Filtering}
%The estimation of the ionospheric scintillation coefficients and the
%production of the corresponding maps based on the observation model
%(eq.~\ref{eq:ObservationModel}) is an inverse problem.
Ionospheric scintillation is a dynamic phenomenon and the quantities
of primary interest are time dependant, thus we will focus on the derivation of a Bayesian scheme for
the
 estimation of the ionospheric scintillation coefficients and the
production of the corresponding images using model
(\ref{eq:ObservationModel}) in a non-stationary statistical
framework under Gaussian assumptions. This dynamic phenomenon
is
expressed %as transformed into
as a state estimation problem \cite{2005Kaipio}
predicted at each time step $t$ using
simple temporal evolution assumptions %on spatial and temporal prior information
and then corrected with the help of scintillation observations
and prior information.
\subsubsection{Evolution model}\label{sec:modeling}
%\subsubsection{Transition probability using smoothness prior}
Since a well-established macroscopic physics-based model, which can explicitly
be described  mathematically and that explains/visualizes the
evolution of the scintillation phenomenon, does not exist, as a first
approach to formulating the time evolution can be
considered a standard %and least-informative
state transition model
which is the random walk model. By using this model, the ionospheric
scintillation is assumed to remain (almost) unchanged between
subsequent steps.

Hence, for the transition between states, we can enforce %the
%simplest kind of random walk
\begin{equation}\label{eq:evolutionModel_walk}
s_t = s_{t-1}+n_t,
\end{equation}
where the state noise $n_t$ can be considered as a variable with a known Gaussian
distribution with zero mean and covariance matrix $\Gamma_{n_t}$
i.e. $n_t\sim\mathcal{N}(0,\Gamma_{n_t})$.
%From Bayes' rule, the forward state transition distribution is
%\begin{equation}\label{eq:Bayesrule_transition}
%\pi(s_t|s_{t-1})\propto \pi_\mathrm{pr}(s_t) \pi(s_{t-1}|s_t),
%\end{equation}
%where $\pi(s_{t-1}|s_t)$ can be seen as a back transition density
%from $s_{t}$ to $s_{t-1}$, and the prior can be Gaussian smoothness
%given by
%Moreover, since we do not have any model to describe the time
Based on (\ref{eq:evolutionModel_walk}), we can infer that the
transition probability density is
\begin{equation}\label{eq:evolutionModel_walk_1}
\pi(s_{t}|s_{t-1}) \propto
\exp{\left(-\frac{1}{2}(s_t-s_{t-1})^\mathrm{T}\Gamma_{n_t}^{-1}(s_t-s_{t-1})\right)}.
\end{equation}

\subsubsection{Augmented observation model} 

The main bottleneck for
the application of the Kalman filter is the limited number of
observations in (\ref{eq:ObservationModel}) (i.e. $M_t<<N$ at every
single instant $t$). This reduces any possibilities for
estimating a time-evolving spatially distributed scintillation index
$s_t$ (i.e. we can estimate images in very limited regions which cannot
be updated continuously since the available measurements cover
different ionospheric areas over time). A significant improvement in the image reconstructions can be achieved by
incorporating a more informative observation model into the Kalman
filtering. Therefore, we propose to include spatial prior information
about the scintillation index and subsequently include it into the
observation model.

Here, we build up a new augmented observation
model. %which allows us to obtain high resolution real time images of the
%ionospheric scintillation using the Kalman filtering.
In particular, let us assume that the scintillation values $s_t$ can
be related using a connectivity matrix $L\in\mathbb{R}^{N\times N}$
\cite{2005Kaipio}, then we can obtain an extra set of equations
\begin{equation}\label{eq:laplace}
L s_t + e=\mathrm{0},
\end{equation}
where $e\in\mathbb{R}^{N}$ is a small perturbation term that follows
a Gaussian distribution given by $e\sim\mathcal{N}(0,\lambda^{-1}
I_{N})$ ($I_N$ is the identity matrix of size $N$) and $\lambda$ a
tuning parameter. Our rationale here is that often
scintillation appears in the ionosphere
as ``clouds'' and thus neighboring ionospheric regions can share almost
similar scintillation activity \cite{Priyadarshi2015}.

From (\ref{eq:laplace}) and (\ref{eq:ObservationModel}),
the augmented observation model is
\begin{equation}
y^{aug}_t = \begin{bmatrix}y_t \\ \mathrm{0} \end{bmatrix} =
\begin{bmatrix}A_t\\ L\end{bmatrix} s_t +\begin{bmatrix}
\varepsilon_t\\ e\end{bmatrix}.
\label{eq:ObservationModelFinal}
\end{equation}
The state-space model is described by
(\ref{eq:evolutionModel_walk}) and (\ref{eq:ObservationModelFinal})
%We note that
where the noise components $n_t$, $\varepsilon_t$ and $e$ are
mutually uncorrelated.
For this observation model, the likelihood $\pi(y^{aug}_t|s_t)$ is 
\begin{equation}\label{eq:likelhood}
%\begin{split}
\pi(y^{aug}_t|s_t) \propto  
\exp{\left(-\frac{1}{2}(y_t-A_ts_t)^\mathrm{T}\Gamma_{\varepsilon_t}^{-1}
(y_t-A_ts_t)-\frac{\lambda}{2}s_t^\mathrm{T}L^{\mathrm{T}}L
s_t\right)}.
%\end{split}
\end{equation}
%The previous probability can be written as a multiplication
%$\pi(y^{aug}_t|s_t) =\pi(y_t|s_t) \pi_{\mathrm{}}(s_t)$.

\subsubsection{Estimates using the proposed state-space model}
In the Bayesian framework, we estimate the marginal posterior distribution of
the state $s_t$ at time $t$ conditioned on the history of the
observations up to the time step $t$ \cite{2005Kaipio}. If we denote
all the observations up to time $t$ as
$$D_t = \{y_1^{aug},y_2^{aug},\ldots,y^{aug}_{t}\},$$ then
the posterior $\pi(s_t|D_t)$ based on Bayes' rule is
\begin{equation}\label{eq:posterior}
\pi(s_t|D_t) \propto
\pi(y^{aug}_t|s_t)\pi(s_t|D_{t-1}),%=\pi(y_t|s_t)\pi_{\mathrm{}}(s_t)\pi(s_t|D_{t-1}),
\end{equation}
where $\pi(y^{aug}_t|s_t)$ is the likelihood (\ref{eq:likelhood})
and $\pi(s_t|D_{t-1})$ is the
 predictive distribution since it predicts $s_t$ at time step $t$ given the measurements up to time
 $t-1$. Based on the Kalman modelling, the
 posterior distribution $\pi(s_t|D_t)$ is Gaussian (denoted for short by $\mathcal{N}(\hat{s}_t,\Gamma_{s_t|D_t})$) with mean  $\hat{s}_t$
 and posterior covariance matrix $\Gamma_{s_t|D_t}$.

The predictive distribution  $\pi(s_t|D_{t-1})$ (see further details in Appendix \ref{Appendix_1}) is
\begin{equation}\label{eq:historyprior}
\pi(s_t|D_{t-1})\propto \exp{\left(-\frac{1}{2}
(s_t-s_t^*)^\mathrm{T}\Gamma_{s_t|D_{t-1}}^{-1} (s_t-s_t^*)
\right)},
\end{equation}
where the predicted vector is
\begin{equation}
  \label{meanandcovariance}
  \begin{gathered}
    s_t^* = \hat{s}_{t-1}
  \end{gathered}
\end{equation}
and the predicted covariance is
\begin{equation}
    \Gamma_{s_t|D_{t-1}} = \Gamma_{n_t}+\Gamma_{s_{t-1}|D_{t-1}},
\end{equation}
with $\Gamma_{s_{t-1}|D_{t-1}}$ being the posterior covariance at
time $t-1$.

By substituting (\ref{eq:likelhood}) and (\ref{eq:historyprior})
into (\ref{eq:posterior}) we have that the marginal posterior $\pi(s_t|D_t)$ has 
mean 
\begin{equation}\label{eq:mean_member}
\hat{s}_t = \Gamma_{s_t|D_t}
(A_t^\mathrm{T}\Gamma_{\varepsilon_t}^{-1} y_t+
\Gamma_{s_t|D_{t-1}}^{-1} s^*_t),
\end{equation}
which is the state estimate at $t$  with
  $\Gamma_{s_t|D_t}$ being the posterior covariance 
\begin{equation}\label{eq:posteriorCovariance_member}
\Gamma_{s_t|D_t} =
(A_t^\mathrm{T}\Gamma_{\varepsilon_t}^{-1}A_t+\lambda
L^\mathrm{T}L+\Gamma_{s_t|D_{t-1}}^{-1})^{-1}.
\end{equation}
%%%%%%%%%%%%%%%%%%%%%%%%%%%%%%%%%%%%%%%%%%%%%%%%%%%%%%%%%%%%%%%
\section{Methodology}\label{sec:material}

\subsection{Single shell ionospheric model and observations}
\begin{figure}[h]
\centering
          \includegraphics[width=0.34\columnwidth]{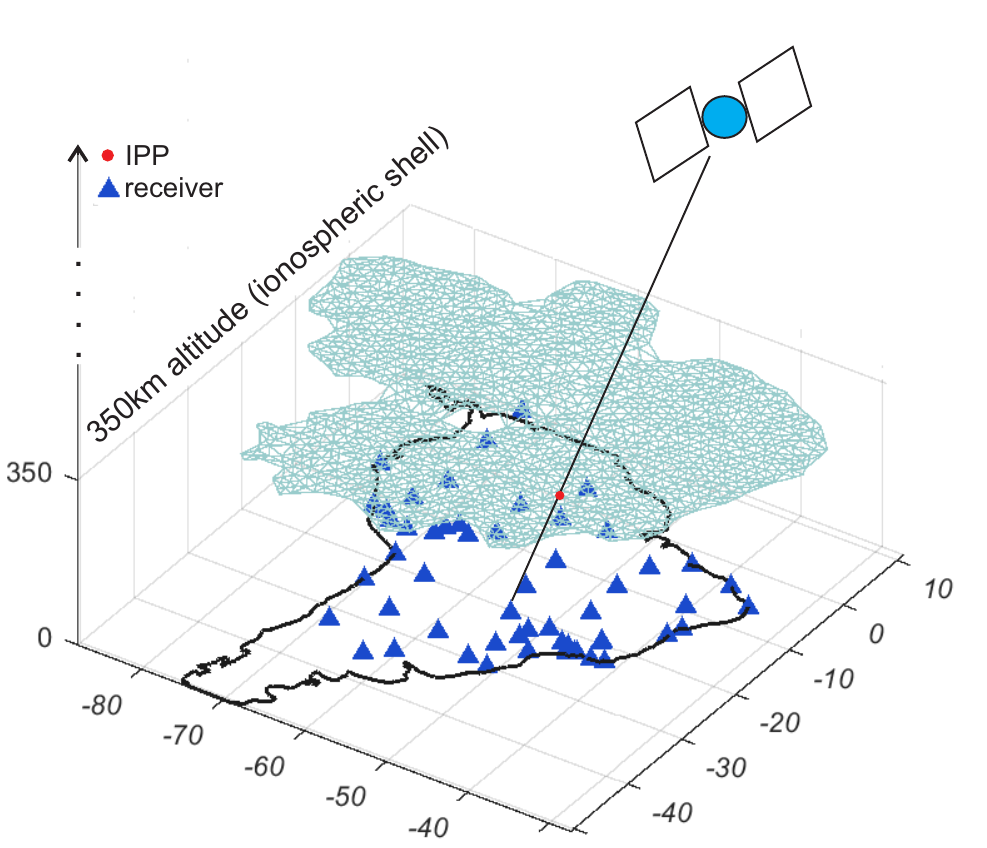}
\caption{Direct projection of the measured data (at a ground receiver) to the ionospheric shell (red dot).}
 \label{fig:project_rule}
\end{figure}

In this paragraph, we describe how a set of ground-based
measurements of scintillation activity can be used in order to
create  images that reflect this activity on the ionosphere (tomography).
Particularly,  at time $t$ we have a set of ground measurements
denoted by $\mathrm{z}_t\in \mathbb{R}^{M_t}$ where $M_t$ is the
number of available links between ground scintillation receivers and
satellites at time $t$.
 %Time index $t$ is defined as
%$t=\lfloor\frac{\bar{t}-\bar{t}_0}{\Delta t}\rfloor$ ,with $\bar{t}$
%the universal time and $\bar{t}_0$ a reference time point.
To construct an ionospheric image, the ground data has to be related
to the ionospheric regions. This is done by projecting the ground
measurements to the ionosphere. If a projector rule is given by
$P:\mathbb{R}^{M_t} \rightarrow \mathbb{R}^{M_t}$, then % has
%to be used, i.e.  %to relate the ground monitors with corresponding
%ionospheric regions. Mathematically, this can be described as
%\begin{equation}
$\mathrm{y}(\mathrm{x}_\mathrm{ion}(t)) = P(\mathrm{z}_t)$,
%\end{equation}
where $\mathrm{y}(\mathrm{x}_\mathrm{ion}(t))\in\mathbb{R}^{M_t}$
are the projected scintillation measurements at the ionospheric
locations $\mathrm{x}_\mathrm{ion}(t)$ which for short are written as
$y_t$ in the following text (see Figure \ref{fig:project_rule}).
By exploiting the single shell model similarly as in \cite{Koulouri2020} and references therein, the ground data at time $t$ can be projected on the
ionospheric layer at the corresponding ionospheric pierce points
(IPPs) at 350 km (considering no angular dependency). Hence, the obtained
observation vector $y_t\in\mathbb{R}^{M_t}$ includes a set of
projected scintillation measurements on the ionospheric shell.
Since all the ground monitors do not simultaneously produce data, 
we note that size $M_t$ varies at each
time step. % (which explains why the subscript $t$ has been included). 

\subsection{Linear observation model}
By exploiting the finite element method (FEM), the ionospheric
region, denoted by $\Omega$, is discretized  and is expressed as
$\Omega \equiv (\mathcal{N},\mathcal{T})$, where
$\mathcal{N}=\{\mathrm{x}_i\}_{i=1:N}$ is the set of nodes and
$\mathcal{T}$ is the set of elements.
Then, the projected scintillation measurements
$y_t[j]$ can be expressed as a linear
combination of basis functions $\phi_i$ with scintillation
coefficients denoted by $s_t[j]$ at each node. Thus, %we can write
\begin{equation}\label{eq:LinearEquationsIonosphere}
y_t[j] = \sum_{i=1}^N\phi_i(\mathrm{x}_j(t))
s_t[j]+\varepsilon_t[j],
\end{equation}
for $j=1,\ldots,M_t$, where $\mathrm{x}_j(t)$ is the ionospheric location of the projected
$y_t[j]$ and $\varepsilon_t[j]$ is the uncertainty introduced due to
numerical approximation and unknown ionospheric disturbances. By
concatenating the linear equations
(\ref{eq:LinearEquationsIonosphere}), we obtain in matrix-form the
limited observation model (\ref{eq:ObservationModel}).  In
particular, in each time step given the projected observations
$y_t[j]$ we estimate the corresponding coefficients $a_t^{ji}$ of
matrix $A_t$ in (\ref{eq:ObservationModel}) where $a_t[j,i]
=\phi_i(\mathrm{x}_j(t))$. Here, we use linear basis functions
$\phi_i$. The support of $\phi_i(\mathrm{x})$ is limited on the
neighborhood of node $i$ (i.e. only on the elements that include
node $i$) and $\phi_i(\mathrm{x}_i) = 1$ at node
$\mathrm{x}_i$ (see Appendix \ref{Appendix_2}). % Thus, for an observation $y_j(t)$ at IPP
%$\mathrm{x}_j$ only the basis functions $\phi_i$ with maximum values
%in the nodes that surround $\mathrm{x}_j$ give nonzero coefficients.

\subsection{Laplace smoothness}
The available instantaneous observations are very limited, thus the
evolution model has been built in such a way that it imposes
connectivity and smoothness between neighboring values which then
are propagated in the next time instant. This consideration is
in-line with the use of the thin shell model which imposes the
projection of a 3D phenomenon into a 2D plane and therefore a 2D ionospheric image reflects the scintillation activity as a result of the superposition of
medium to small-scale % (vertical and horizontal) 
plasma irregularities \cite{Yeh2020}.
%fluctuating plasma structures.
%nd the fact that an ionospheric map reflectsthe  superposition  of  medium  to  small-scale  spatial  structuresrelated to scintillation

In the current implementation, matrix $L$ in
(\ref{eq:ObservationModelFinal}), is %selected to be 
the 
normalized Graph Laplace \cite{Murphy2012}, given by
\begin{equation}
L=\mathrm{I}_N-D^{-1} H,
\label{eq:NormLaplace}
\end{equation}
where $\mathrm{I}_{N}$ is the identity matrix of size $N$;
$D\in\mathbb{R}^{N\times N}$ is a diagonal matrix with diagonal
elements $D_{ii}=\sum_{j=1}^N H_{ij}$; and matrix
$H\in\mathbb{R}^{N\times N}$ has non zero elements for $i\neq j$,
$H_{ij}= - \frac{1}{h_{ij}}$, if nodes
$i$ and $j$ are connected (e.g. with a vertex) otherwise $H_{ij}=0$ and $H_{ii}=0$ . % and
%the diagonal element equal to $H_{ii}=-\sum_j h_{ij}$
Here, $h_{ij}$ is the longitude/latitude distance between the
ionospheric nodes $i$ and $j$. This choice of Laplace ensures a smooth transition between neighboring nodes (in the area where measurements exist) and a uniform value for the $s_t$ over the area where there are not any or very low influence from the available observations (since the the null space of $L$ is the unit vector). 
\subsection{Initial conditions}
For the observation model (\ref{eq:ObservationModelFinal}), the
noise term $\varepsilon_t$ was modelled i.i.d. Gaussian with
covariance $\Gamma_{\varepsilon_t}=\gamma_\varepsilon I_{M_t}$. This
error corresponds to the uncertainties introduced due to the
ionospheric projection, inherent measurement errors and unknown
ionospheric disturbances. In the results section \ref{section:results}, the used value was
$\gamma_\varepsilon=0.0018$, this value corresponds to the variance
of the errors between observations $y_t[j]$ at neighboring IPPs
$\mathrm{x}_j(t)$ (less than 1km distance) given time $t$, i.e. it
was estimated from  differences of measurements,  $y_{t_k}[l] -
y_{t_k}[j]$ when their corresponding IPPs satisfied
$\|\mathrm{x}_{l}(t_k)-\mathrm{x}_{j}(t_k)\|_2\leq
1\;\mathrm{km}$ (in  Cartesian coordinates). For the evolution model
(\ref{eq:evolutionModel_walk}), the covariance of the perturbation
$N_t$ was modelled as $\Gamma_{n_t}=\gamma_n I_N$.
Here, $\gamma_n$ was set equal to the variance of the difference of
consecutive observations corresponding to the same satellite-ground
monitor link.
Finally, following the Kalman filter initial condition $\pi(s_0|D_0)
=\pi(s_0) $, we set $\Gamma_{s_1|D_0}=\Gamma_{n_t}$ and
$\hat{s}_0=0$.

\subsection{Ensemble of Kalman filters}

\begin{figure}[h]
 \centering
          \includegraphics[width=0.5\columnwidth]{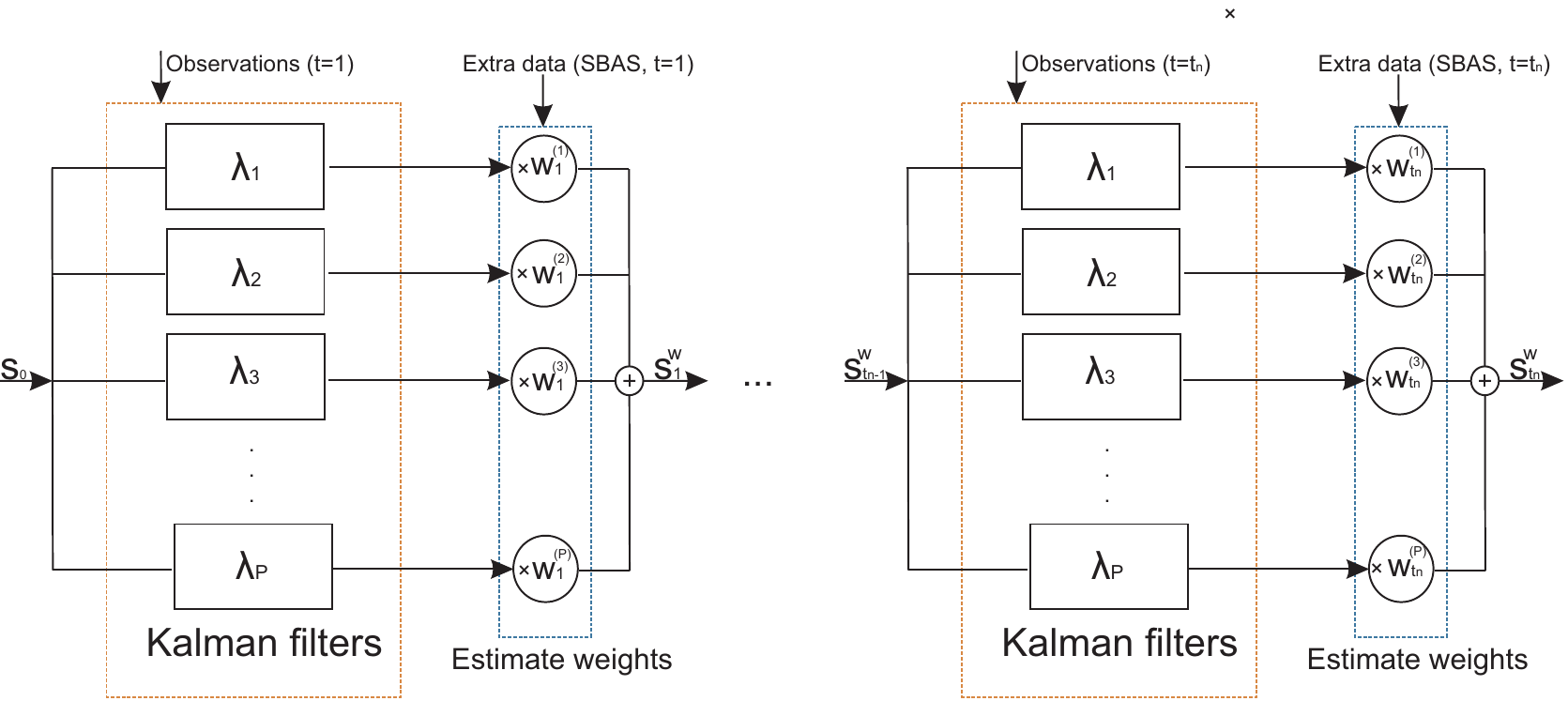}
 \caption{Pipeline of the proposed ensemble of Kalman filters. %As we will see next, the control data is recorded from monitors that use the SBAS constellation.
 }
 \label{fig:Kalman}
\end{figure}

The selection of a tuning parameter $\lambda$ is not a
trivial task particularly in a time varying problem.  Here, we employ
the multiprocess modelling concept \cite{Murphy2012,Magill1965} and
we introduce an ensemble of $P$ state-space models
((\ref{eq:evolutionModel_walk}) and
(\ref{eq:ObservationModelFinal})), each one indexed by an indicator
$p$ with a tuning parameter $\lambda$ taken from a discrete
set of values $\lambda_p$ for $p =1,\ldots,P$ (where $P$ is the total
number of tuning parameters). The basic idea of using an
ensemble of state-space models is that a priori no constant
tuning parameter $\lambda$ is expected to hold for all time
instances $t$.
Therefore, by obtaining estimates at each time $t$ using different
tuning parameters and then estimating a weighted
average (with weights relying on control data as we shall see next),
we can overcome the $\lambda$ selection problem. Figure \ref{fig:Kalman} illustrates the proposed pipeline.  %at each time step. 

In particular, the moments (mean and covariance) of  each member
of the ensemble are obtained by
running in parallel $P$  Kalman filters, each conditional on %assuming
%that
%Bayesian Moment Matching, we project the posterior onto a tractable family of distribution by matching a set of sufficient moments. 
the state of $\lambda=\lambda_p$. Then, an ionospheric scintillation
image is estimated at $t$ as a weighted average
\begin{equation}\label{eq:mean}
\hat{s}^w_t = \sum_{p=1}^P w_t^{(p)} \hat{s}_t^{(p)},
\end{equation}
where $\hat{s}_t^{(p)}\in \mathbb{R}^N$ is the  mean of the $p^{th}$ member given $\lambda_p$
and $w_t^{(p)}>0$ are weights, showing our confidence on the
selected $\lambda_p$.
The weights can be estimated on-line using the residuals
between available held-out data and predicted estimates obtained from %the
%outputs of 
the $P$ Kalman filters. %, as described in the results
%section. 
Furthermore, the uncertainty in the estimates (\ref{eq:mean}) can be
quantified with the weighted covariance matrix
\begin{equation}
\begin{split}
{\Gamma}_{t}^w=&\mathbb{E}[(s_{t}-\hat{s}^w_t)(s_{t}-\hat{s}^w_{t})^\mathrm{T}]\\=&
\sum_{p=1}^P w_{t}^{(p)}
(\Gamma_{s_{t}|D_{t}}^{(p)}+(\hat{s}_{t}^{(p)}-\hat{s}^w_{t})(\hat{s}_{t}^{(p)}-\hat{s}^w_{t})^\mathrm{T}).
\end{split}
\end{equation}
Then, the ensemble of Kalman filters is described in Algorithm 1. 
Based on Algorithm 1, the weighted mean and covariance are used at the prediction step of the ensemble members. In effect here we have a hybrid model that expresses our belief state with a mixture of $P$ Gaussians \cite{Alspach1972} which is approximated as a single Gaussian \cite{Maybeck1979,Opper1999}. 

\begin{algorithm*}[t]%tpt]
\caption{Sequence of ionospheric scintillation images using parallel Kalman filters} \label{alg:KalmanScintillation}
    \begin{algorithmic}
     \State \textbf{Initialization}: $\hat{s}^w_0=0,\;s_0^{(p)}=0 \in \mathbb{R}^N$, $\{\lambda\}_{p=1:P}$, $\Gamma_{n_t}=\gamma_n I_{N}$,
   $\Gamma_{\varepsilon_t}=\gamma_\varepsilon I_{M_t}$ and
   $\Gamma_0^w =0$.
  \For{$t = 1...$} \Comment{for every minute}
         \State  Estimate the nonzero coefficients of matrix $A_t$ using the available data $y_t$;
           \For{$p=1:P$} \Comment{run parallel Kalman filters}
         \State  Set $s_t^{(p)*} =\hat{s}^w_t$ and
         $\Gamma_{s_t|D_{t-1}}^{(p)}=\Gamma_{n_t}+\Gamma_{t}^w$
         \Comment{prediction step}
       % \State Estimate prediction errors %$r_t^{(p)*}=[A_t\; L]^\mathrm{T} s_t^{(p)*}-[y_t\;0]^\mathrm{T}$
        %$r_t^{(p)*}=A_t s_t^{(p)*}-y_t^\mathrm{T}$ and covariance (\ref{eq:predictedResidual})
       % and $\alpha^{(p)}_t$ parameter based on (\ref{alpha_tunning})
        \State  Estimate member's covariance: $\Gamma^{(p)}_{s_t|D_t} = (A_t^\mathrm{T}\Gamma_{\varepsilon_t}^{-1}A_t+\lambda_pL^\mathrm{T}L+ \Gamma_{s_t|D_{t-1}}^{(p)-1})^{-1}$
        \State  Estimate member's mean: $ \hat{s}^{p}_t =
        \Gamma_{s_t|D_t}^{(p)} (A_t^\mathrm{T}\Gamma_{\varepsilon_t}^{-1} y_t+ \Gamma_{s_t|D_{t-1}}^{(p)-1} s^{(p)*}_t)$  \Comment{updates}
        \EndFor
        \State  Estimate the weights $w_t^{(p)}$ as described in
        section~\ref{sec:weights} using control data $y^c(t)$.
        \State Estimate scintillation image: $\hat{s}_t^{w}=\sum_{p=1}^Pw_t^{(p)} \hat{s}^{(p)}_t$
        \State Estimate covariance: ${\Gamma}_{t}^w=\sum_{p=1}^P
w_{t}^{(p)}
(\Gamma_{s_{t}|D_{t}}^{(p)}+(\hat{s}_{t}^{(p)}-\hat{s}^w_t)(\hat{s}_{t}^{(p)}-\hat{s}^w_t)^\mathrm{T})$
 \EndFor
    \end{algorithmic}
\end{algorithm*}
%\end{multicols}{2}
%\end{figure*}
 
\subsubsection{Estimation of the weights}\label{sec:weights}
From Bayes' rule \cite{Chaer1997,2005Kaipio}, the weights
in~(\ref{eq:mean}) can be expressed as probability densities
$w_t^{(p)}=\pi(\lambda_p|y_{t}^c)$  where
\begin{equation}
\pi(\lambda_p|y_{t}^c) = \frac{\pi(y_{t}^c|\lambda_p)
\pi(\lambda_p)}{\sum_{p=1}^P \pi(y_{t}^c|\lambda_p) \pi(\lambda_p)}.
\end{equation}
Here, $y_{t}^c\in\mathbb{R}^{M_{c_t}}$ denotes a set of
control data of size $M_{c_t}$ that was not used to
estimate each (member's) mean $\hat{s}^{(p)}_t$, and covariance
$\Gamma^{(p)}_{s_t|D_t}$.

Without prior information  about the value of the tuning
parameter, we can consider that $\lambda_p$ are uniformly
distributed i.e. $\pi(\lambda_p)=1/P$. Thus,
\begin{equation}
w_t^{(p)} = \frac{\pi(y_{t}^c|\lambda_p)}{\sum_{p=1}^P
\pi(y_{t}^c|\lambda_p)}.
\end{equation}
The conditional densities are modelled as Laplace distributions
\begin{equation}
\pi(y_{t}^c|\lambda_p)\propto
\exp{\left(-\sum_{m=1}^{M^c_t}\kappa_m|y_t^c[m]-\hat{y}^{(p)}_t[m]|\right)},\end{equation}
with
$\kappa_m=\left(\sqrt{\frac{1}{2P}\sum_{p=1}^P(y_t^c[m]-\hat{y}^{(p)}_t[m])^2}\right)^{-1}$
and $\hat{y}^{(p)}_t$ being the member prediction estimated from
$\hat{y}^{(p)}_t=\hat{A}_t \hat{s}_t^{(p)}$, where $\hat{s}_t^{(p)}$
is the mean of the $p^{th}$ member and where $\hat{A}_t$ includes the
basis coefficients for the IPP  of  measurement $y_{t}^c$. Here
$[m]$ denotes the $m^{th}$ entry of vector $y_t^c$ and
$\hat{y}^{(p)}_t$. Therefore,
the weights are estimated based on the corresponding residuals
(between the prediction and control data). We note that the
heavier tails of the Laplace distribution (compared to e.g. a
Gaussian distribution which would be the alternative standard
option) can
handle better extreme cases, where control data does not fully agree
with the predictions. % of the ensemble. 

%\vspace{-1.4em}
\clearpage

\section{Data and dynamic scintillation imaging}\label{section:results}

In this section, we describe the available data and present dynamic ionospheric scintillation images in the
area of South America at geographic latitude between $\sim
-40^\circ\mathrm{N}$ and $\sim 10^\circ\mathrm{N}$ and longitude between
$\sim -90^\circ\mathrm{E}$ and $\sim -35^\circ \mathrm{E}$. These
regions experience the most significant amplitude scintillation activity %including deep
%signal fades that can cause a GNSS receiver
%to lose track of one or more satellite signals 
mostly after sunset until a few hours after midnight local time. We believe that creating real time ionospheric images at 350 km
estimated using $S_4$ data in this area is rather important. %, available at 1 minute intervals.

\subsection{Available data}
The $S_4$ data (that was used to produce ionospheric images) was
measured from a network of 36 ISMR scintillation receivers
using all the available satellite systems i.e. the Global
Positioning System (GPS), Globalnaya Navigazionnaya Sputnikovaya
Sistema, (GLONASS), Galileo (European navigation system) and
Satellite-based Augmentation System (SBAS).
In particular, the $S_4$ (L1) measurements (with sampling period $t=1$
minute), the links (satellites-station names) and their \emph{IPPs} at 350 km were downloaded
from the CIGALA/CALIBRA network - UNESP web server \cite{Vani2017,Data} (we note that the downloaded data was used as such without applying any post-processing step). Each instant image was produced using as $y_t$ observations, the $S_4$ measurements
available within that specific minute.
The period that the images were produced was during the night time between 01 and 02 December
2014, a period of time which
was characterized by mild to strong scintillation according to the
information provided by the ISMR Query Tool
(\href{http:// is-cigala-calibra.fct.unesp.br}{http://
is-cigala-calibra.fct.unesp.br}) \cite{Vani2017}. % We note that for the scintillation activity we considered the characterization levels:  
%weak $(0.3\leq S_4<0.4)$, moderate $(0.4\leq S_4<0.7)$ and strong
%$(S4\geq0.7)$ scintillation \cite{Veettil2020}.

\subsubsection{Control data for the weights}
\begin{figure}[h]
 \centering
   \includegraphics[width=0.3\columnwidth]{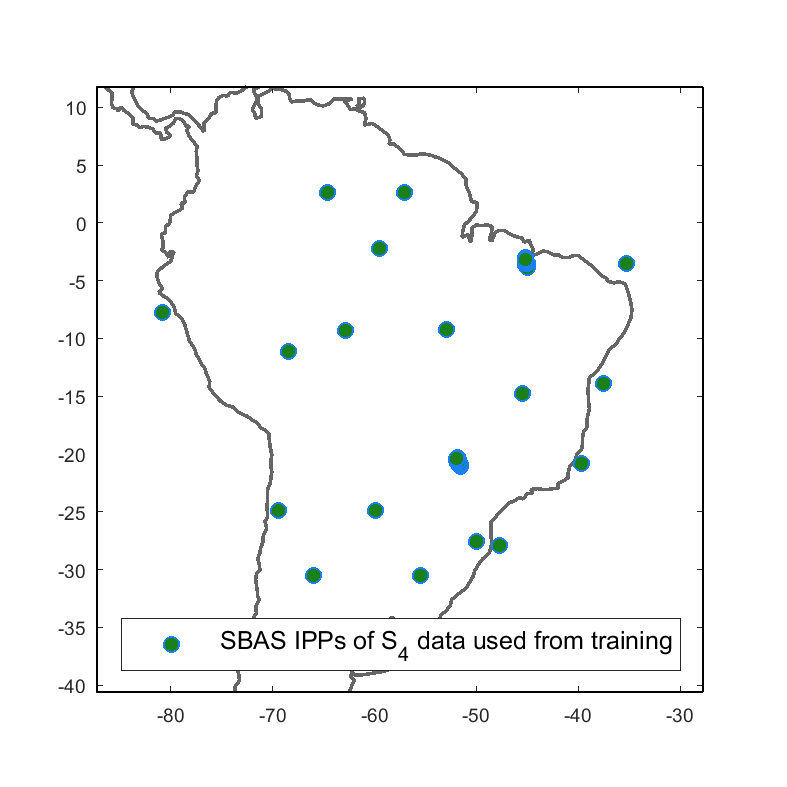}
   \caption{The IPPs of the SBAS control data.}
    \label{fig:training}
   \end{figure}

Approximately $10\%$ of the remaining $S_4$ data, which was available at every minute and corresponded to 19 real-time measurements from
the SBAS system, were used to estimate the weights in the ensemble of
Kalman filters. The SBAS data was chosen for the weighting because it was
continuously available for the same set of (ground receiver-SBAS satellite) links with fixed IPP locations on the ionospheric shell. 
The locations of the IPPs of these 19 measurements appear in
Figure~\ref{fig:training}.

\subsection{Image construction over South America}\label{sec:CompDomainDataAcc}
 \begin{figure}[h]
 \centering
 \begin{tabular}{cc}
           \includegraphics[width=0.2\columnwidth]{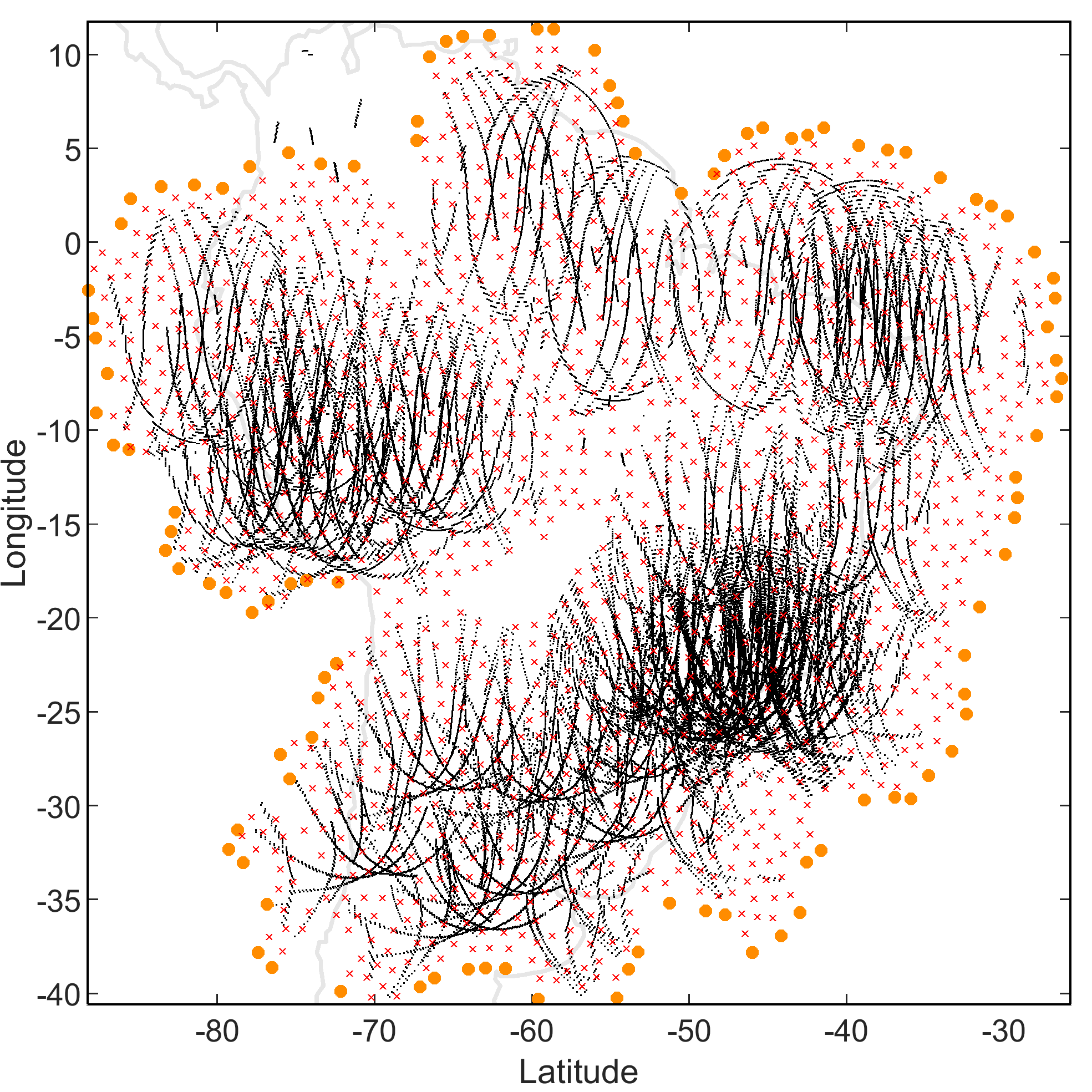}%Links_4Sat_14.png}
           &\includegraphics[width=0.2\columnwidth]{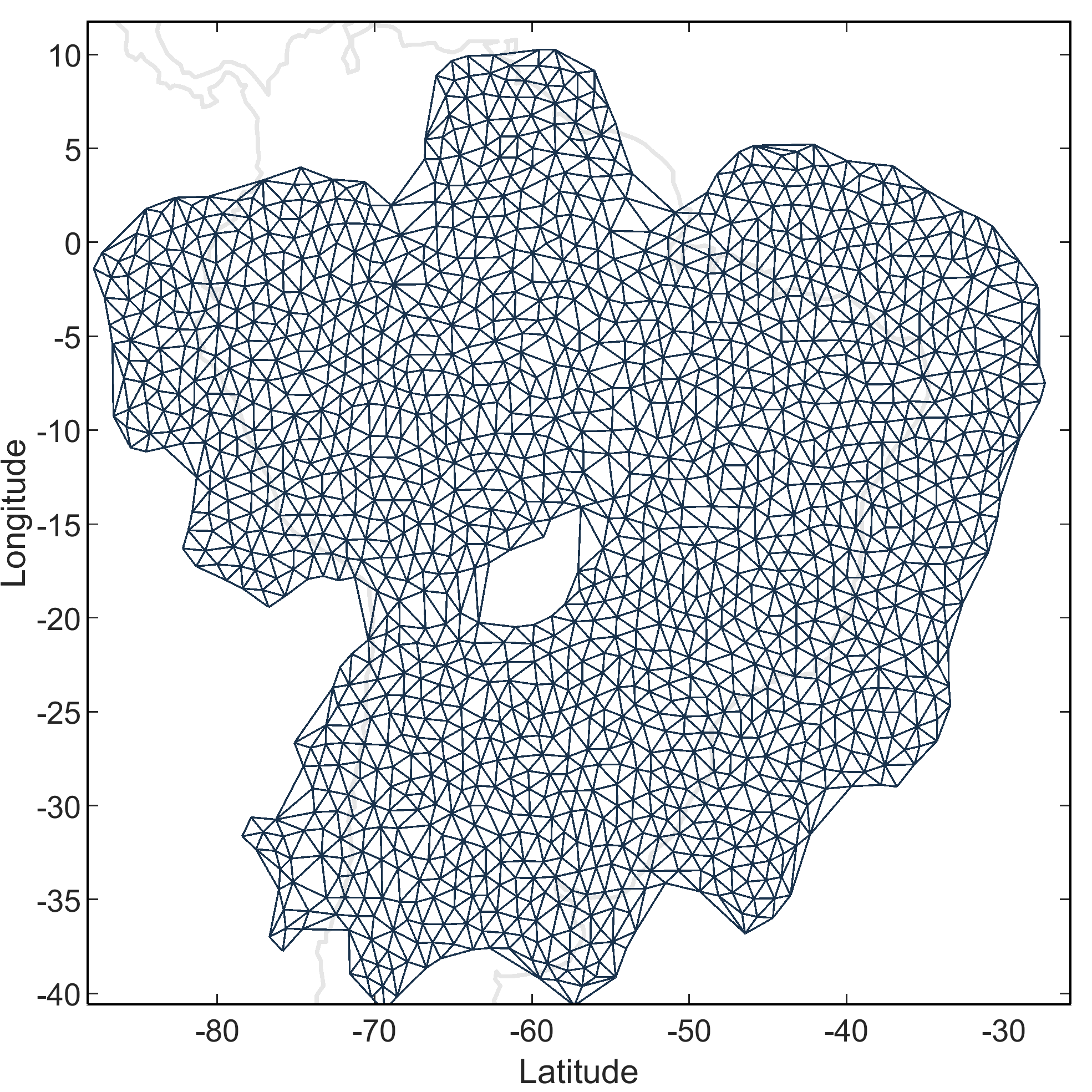}%Links_4Sat_53.png}
           %&\includegraphics[width=0.3\columnwidth]{figures/Links_58.png}
 %\\ (a) & (b)% & (c)
 \end{tabular}
 \caption{Left: Distribution of IPPs over 6 hours (and corresponding borders marked in orange color and nodes in red crosses). Right: Constructed mesh used in the estimation of ionospheric scintillation images.}
\label{figure_gr3}
\end{figure}
Instead of a regular grid which restricts the ability for
adaptations around the locations where there is constant
information, we used a triangular mesh to construct the
ionospheric scintillation images where we defined smaller and larger
elements based on the availability of the observations in the
different ionospheric locations. The number of the nodes of the mesh
and their locations was based on the availability of the projected
$S_4$ measurements over the time interval that the imagess were produced.

To estimate the mesh,
we performed the following steps.
  We plotted all the available IPPs over a period of time
  (i.e. 6 hours) in order to identify the region where there is
  available data.
   Then, we marked all the IPPs that lied on the boundaries of the
  ionospheric area to define the borders of the mesh.
  Subsequently, we down-sampled the IPPs using a circle of radius $R$, i.e.
  we started from one boundary IPP, we discarded all the IPPs within
  a circle of radius $R=1$ (in units of longitude and latitude) with center this IPP, and we kept the IPP  which was the closest to the current one and outside of the circle.
  Then, we used this new point in a similar fashion to
  estimate the next point. The downsampled IPPs were used as the nodes of the created Delaunay   triangular mesh (also the long deformed elements created by boundary nodes were discarded). To avoid very small elements on the boundary, we applied the Taubin's mesh
  smoothing \cite{Taubina}. 
  The nodes of the constructed mesh were the locations where the values
 $s^w_t$ (\ref{eq:mean}) were estimated.
  Figure~\ref{figure_gr3} illustrates on the left the original distribution of
  IPPs, the border marked with orange color, the created nodes in read color and on the right we have the final mesh.

% \begin{figure}[htb]
% \centering
% %\begin{tabular}{cc}
% \includegraphics[width=0.95\columnwidth]{figures/Estimate_MESH.eps}
% % \end{tabular}
%  \vspace{-1.0em}
% \caption{(A): Distribution of IPPs over 24 hours (and corresponding borders marked in orange color). (B): Constructed mesh used in the estimation of ionospheric scintillation maps.}
% \label{figure_gr3}
% \end{figure}

\FloatBarrier

\newpage

\subsection{Real-time scintillation imaging}
\begin{figure}[h]
 \centering
\includegraphics[width=0.5\columnwidth]{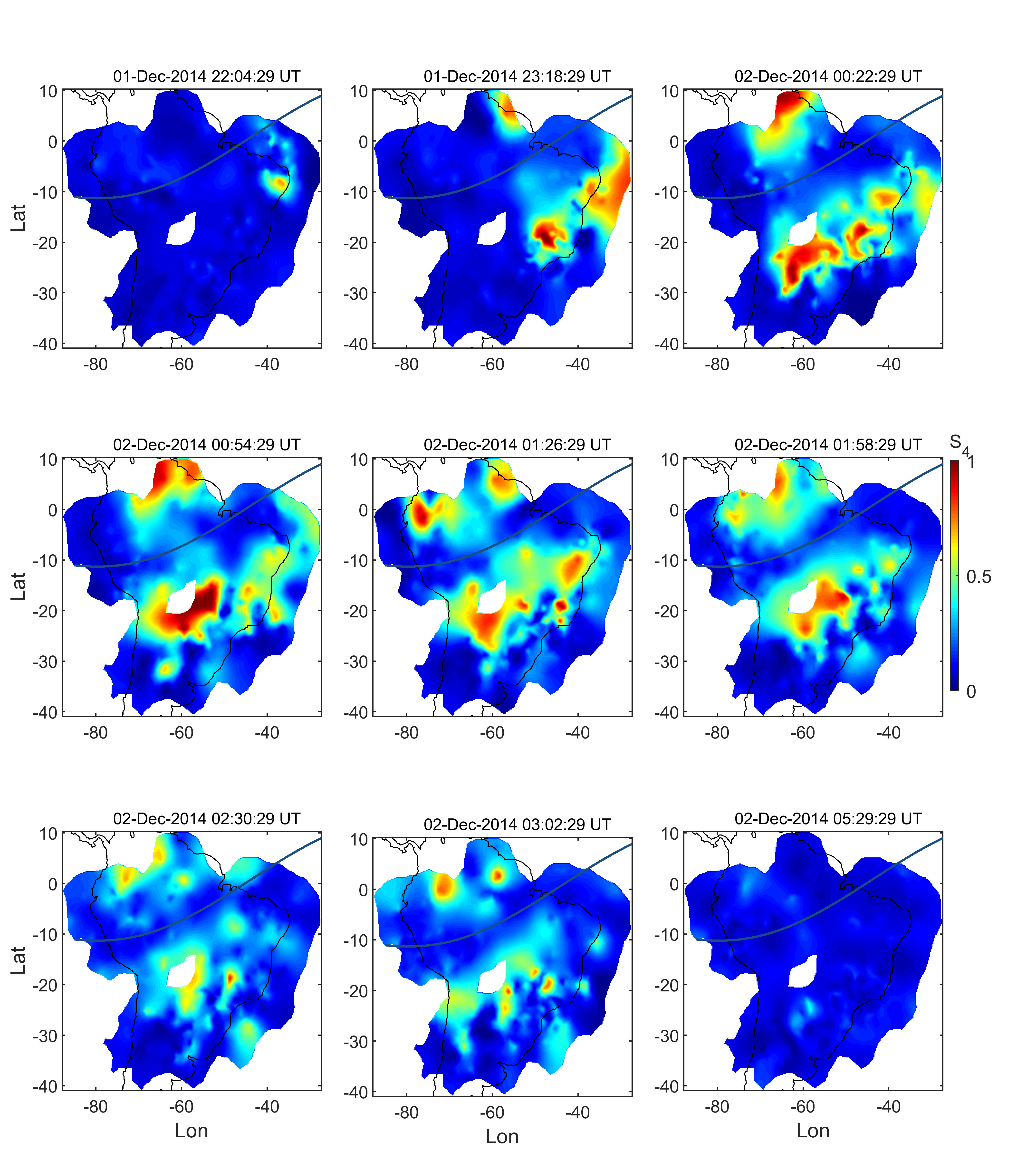}
  \caption{Sequence of $S_4$ ionospheric images (snapshots) during the night-time between December 1 and 2, 2014. The dark line indicates the magnetic equator.}
 \label{figure_grday3}
\end{figure}

To create the ensemble of Kalman filters, we employed 13 tuning values $\lambda_p$ ranging from $10^{-3}$ to $500$. This range was selected in order to avoid over-smoothing the covariance (\ref{eq:posteriorCovariance_member}). In Figure~\ref{figure_grday3}, we can see instantaneous ionospheric scintillation images over South America during the night-time between December 1 and 2, 2014. Each image depicts the estimated $S_4$ distribution on the mesh, employing only the $S_4$ measurements that were available at a minute interval provided above each picture in Universal Time (UT). The blank space indicates an area where there was no coverage from the current GNSS system. 
As we can see the highest scintillation activity is observed over the equatorial anomaly and especially over Brazil. The activity is low early in the evening while it increases during the beginning of the night over each region of the continent (we could see that by converting the UT to local times).
These observations are in line with the static risk maps presented in our previous work \cite{Koulouri2020}. %Add here the link for the video!
%We have to pinpoint as we will see also in the following section that this is also related to the better coverage in that region (more ground stations)
 % \vspace{-1.3em}

\FloatBarrier

\section{Cross-validation tests and analysis} \label{section:Valid}
To validate the accuracy of our modelling, we compare sequences of
real measurements (obtained from given satellite-receiver links) with
predictions estimated using the produced images.
In particular, we used the cross-validation technique i.e. we produce
a sequence of ionospheric maps $\hat{s}_t$ without including the
scintillation data from a test satellite-receiver link (a.ka.left-out
link). Then, we use the ionospheric image and the given IPP for
the left-out link to estimate the scintillation values for this
particular link over time and then we compare our predictions $\hat{y}_t$, 
with the measurements $y^{\mathrm{r}}_t$ (validation data). The same was repeated for
different satellite-monitor links as we can see in this section.
The sequence of the predicted values for a link $k$ was estimated by
$\hat{y}_t[k] =
\sum_{i=1}^N\phi_i(\mathrm{x}_k^\mathrm{r}(t))\hat{s}^w_t[i]=A_t^{k,:}\hat{s}^w_t$,
where $A_t^{k,:}\in\mathbb{R}^{1\times N}$ is a row vector that includes
the basis coefficients for a given IPP $\mathrm{x}_k^\mathrm{r}(t)$, and $\hat{s}^w_t$ is the estimated weighted
ionospheric image at time $t$. The numerical standard deviation was estimated as 
$\sigma_t[k]=\sqrt{A_t^{k,:} [\sum_{p=1}^P w_{t}^{(p)}(\hat{s}_{t}^{(p)}-\hat{s}^w_{t})(\hat{s}_{t}^{(p)}-\hat{s}^w_{t})^\mathrm{T}] (A_t^{k,:})^\mathrm{T}}$.
%sqrt(A_t*Gamma_w_num*A_t')

For the numerical comparison, we employed two metrics: 
the data-model correlation metric (CM)
\begin{equation}
C_{\hat{y},y^\mathrm{r}} =
\frac{1}{T}\frac{1}{\sqrt{\gamma_{\hat{y}}},\sqrt{\gamma_{{y^\mathrm{r}}}}}\sum_{t=0}^T\left(\hat{y}_t-\langle\hat{y}\rangle\right)\left({y}^\mathrm{r}_t-\langle{y}\rangle\right),
\label{eq:CM}
\end{equation}
where $\langle.\rangle$ denotes the mean values and $\gamma$ are the
sample variances of the estimates $\hat{y}$ and measured $S_4$ values  $y^\mathrm{r}$ and 
%  \item 
the Root-mean-square error (RMS)
\begin{equation}
e_\mathrm{rms}=\sqrt{\frac{1}{T}\sum_{t=1}^T(\hat{y}_t-y^\mathrm{r}_t)^2},
\label{eq:RMS}
 \end{equation} 
which gives the difference between the model estimation
and the actual system activity.

\begin{figure}[h]
 \centering
 \begin{tabular}{cc}
           \includegraphics[width=0.2\columnwidth]{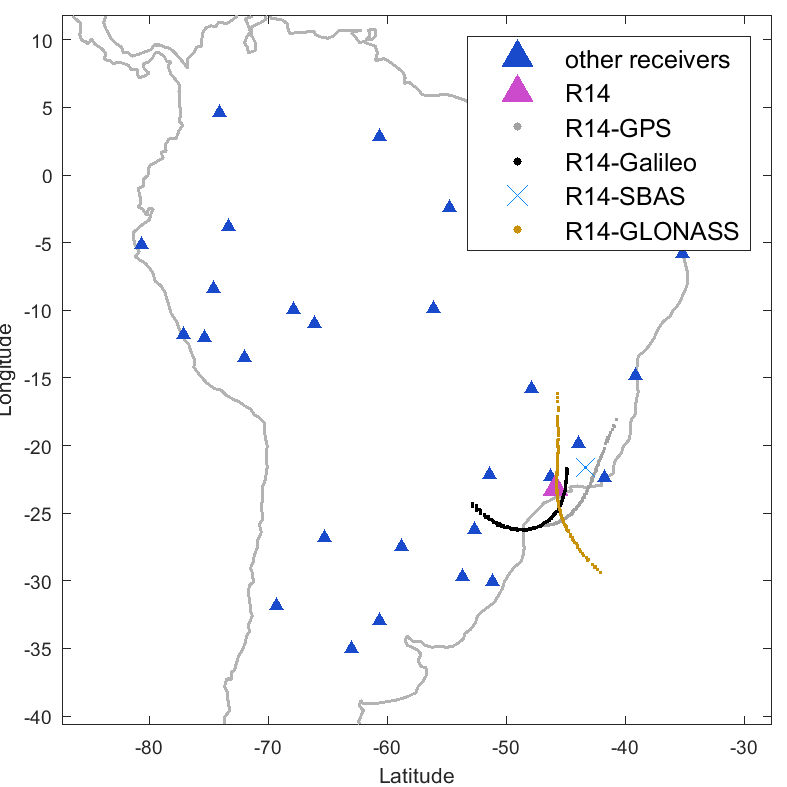}%Links_4Sat_14.png}
           &\includegraphics[width=0.2\columnwidth]{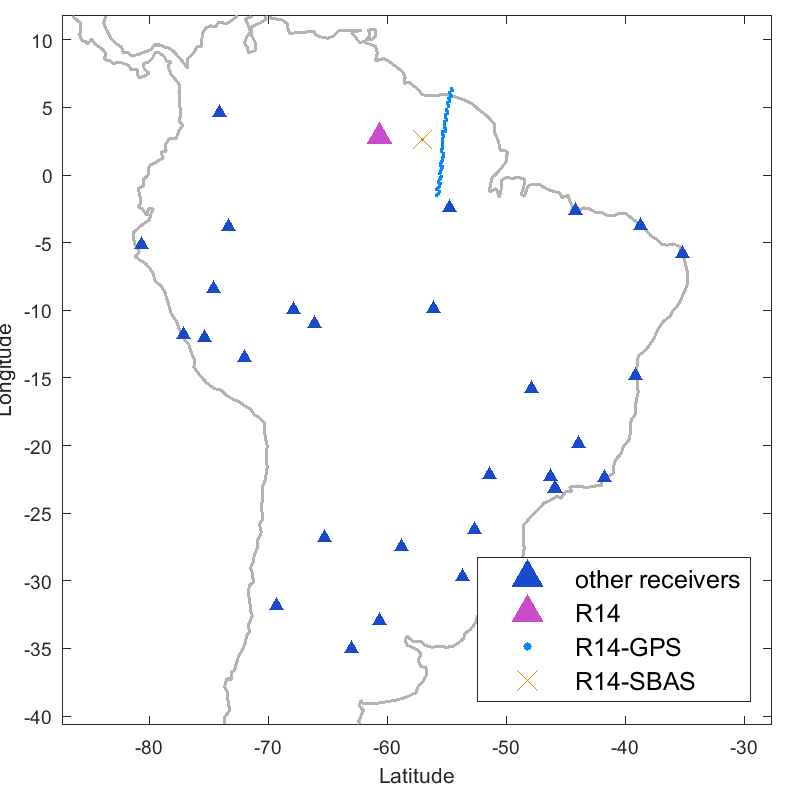}%Links_4Sat_53.png}
           %&\includegraphics[width=0.3\columnwidth]{figures/Links_58.png}
 %\\ (a) & (b)% & (c)
 \end{tabular}

 \caption{Left: Location of receiver 14 and the corresponding IPPs at 350km between 14 and satellites S120 (SBAS), G29 (GPS), R6 (GLONASS), E12 (Galileo). Right: Location of receiver 53  and  the corresponding IPPs between 53 and S120 (SBAS) and G29 (GPS).}%
 \label{figure:Locations}
\end{figure}

\begin{figure}[h]
\centering
\begin{tabular}{cc}
    \includegraphics[width=0.2\columnwidth]{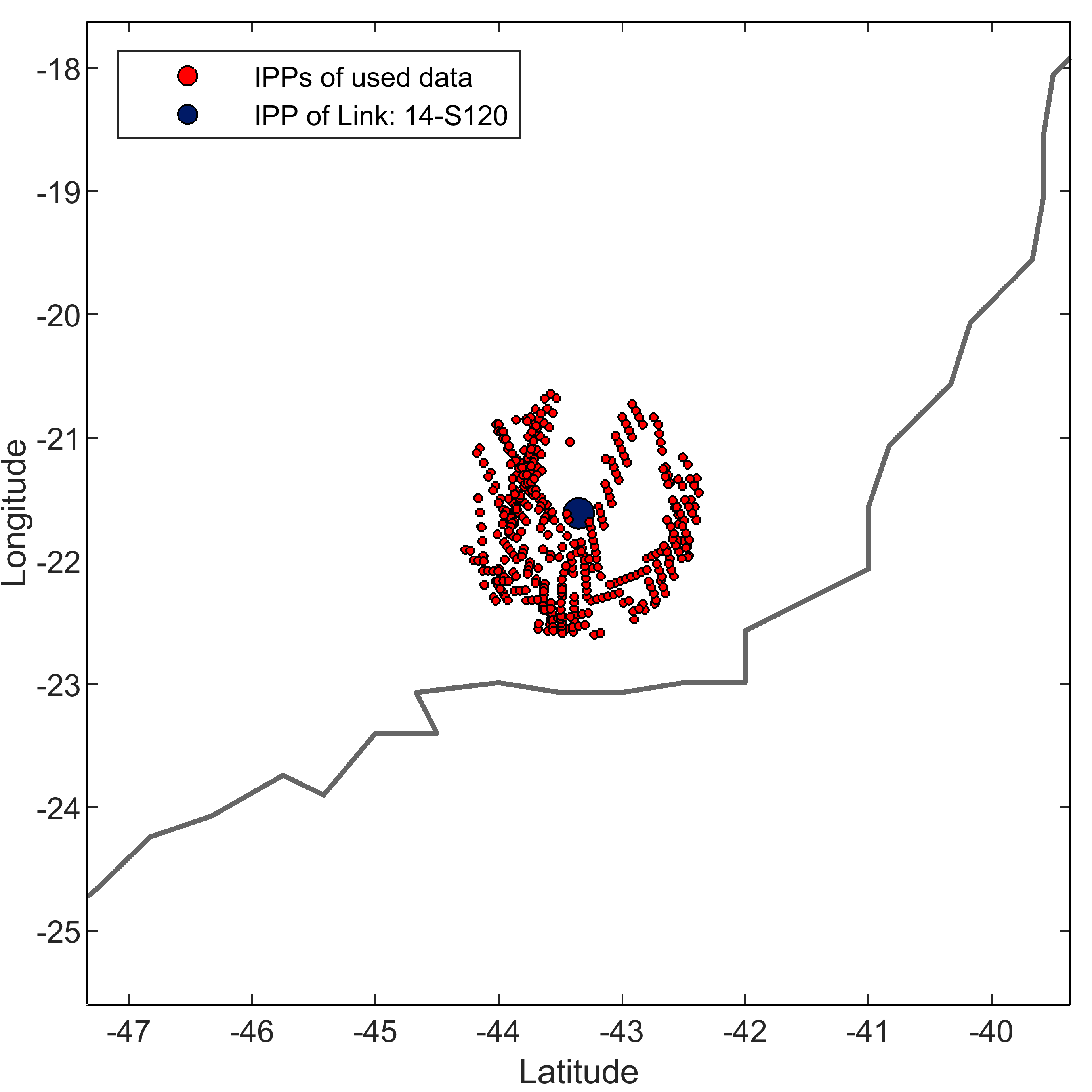}
    & \includegraphics[width=0.2\columnwidth]{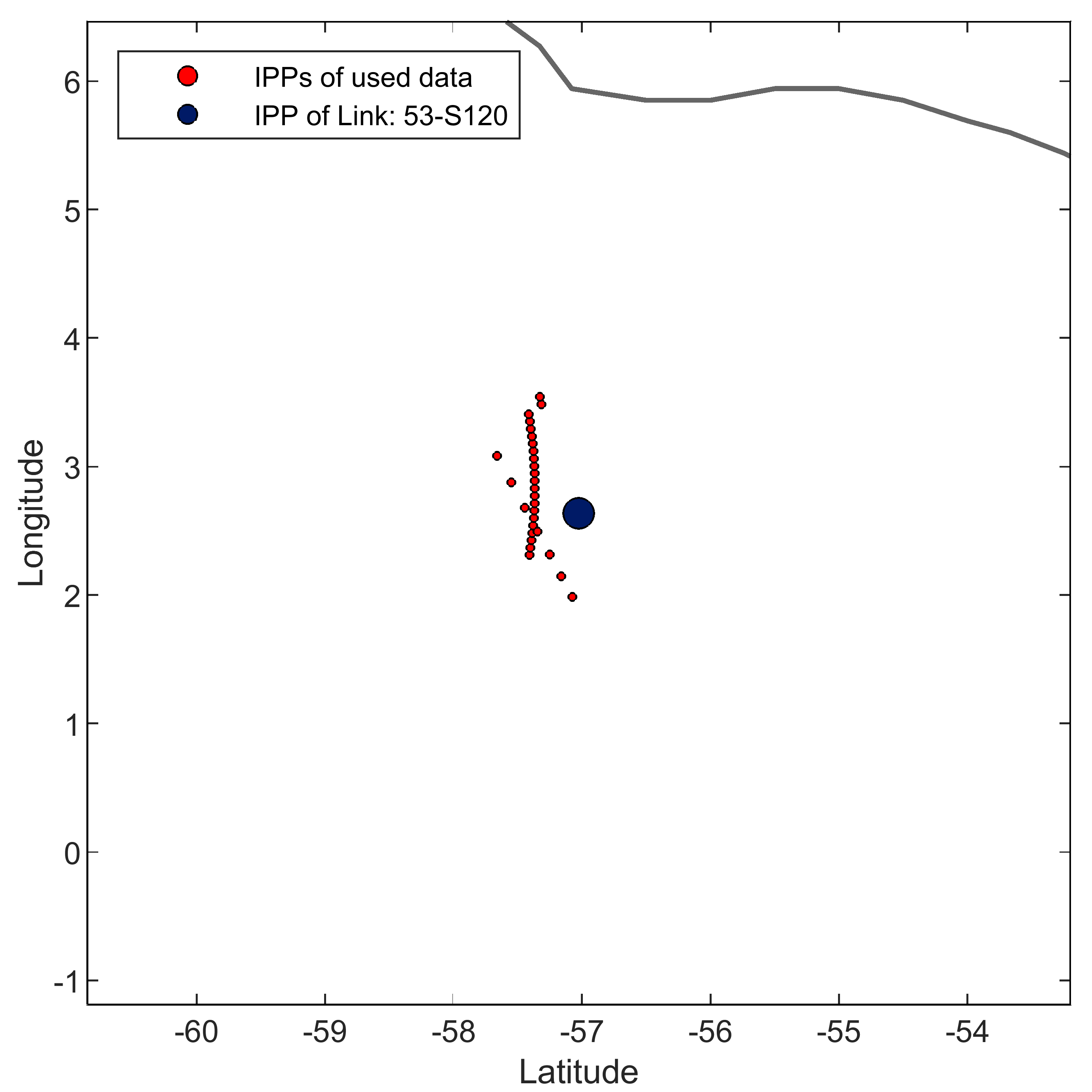}
%    \\ (a) & (b)
\end{tabular}

\caption{%Big images: 
The small dots show the IPPs of the available observations that were used to produce the image (in a circle of radius one) around the IPPs of link R14-S120 (left image) and link R53-S120 (right image).% Small images: The whole continent with the locations of R14 (left) and R53 (right) circled.
}
\label{figure_used_data_locations}
\end{figure}

In the following, we compare our predictions (from the created images) against the measured $S_4$ values from two ground receivers 14 and 53  that were not used in the image estimation (the numbering of the receivers is based on the receiver database available in \cite{Vani2017}). The locations of these receivers are shown in Figure \ref{figure:Locations}. The receiver 14 was chosen because it is surrounded by other nearby ground receivers, and receiver 53 because it is located remotely i.e. there are less ground receivers nearby. As a result, the effect of the spatial density of available observations on the accuracy of the estimated images can be showcased. Figure \ref{figure_used_data_locations}
shows the IPPs of the available observations that were used to produce the predicted $S_4$ values. 

%\begin{figure}[htb]
%\centering
%\includegraphics[width=.6\columnwidth]{figures/Spatial_locations_53_83120with_observations.png}
%\caption{Big image: The small dots show the IPPs of the available observations that were used to produce the instant maps (in a circle of radius two) around the IPP of Link: 53-S120 (denoted with a deep blue circle). Small image: The whole continent with the location of the IPP of Link: 53-S120. % and the small circle with the used data.
%}
%\label{figure_used_data_locations_53}
%\end{figure}
%
%  \vspace{-1.1em}
\FloatBarrier
\subsection{Receiver 14}
\begin{figure}[h]
 \centering
 \begin{tabular}{c}
            \includegraphics[width=0.4\columnwidth]{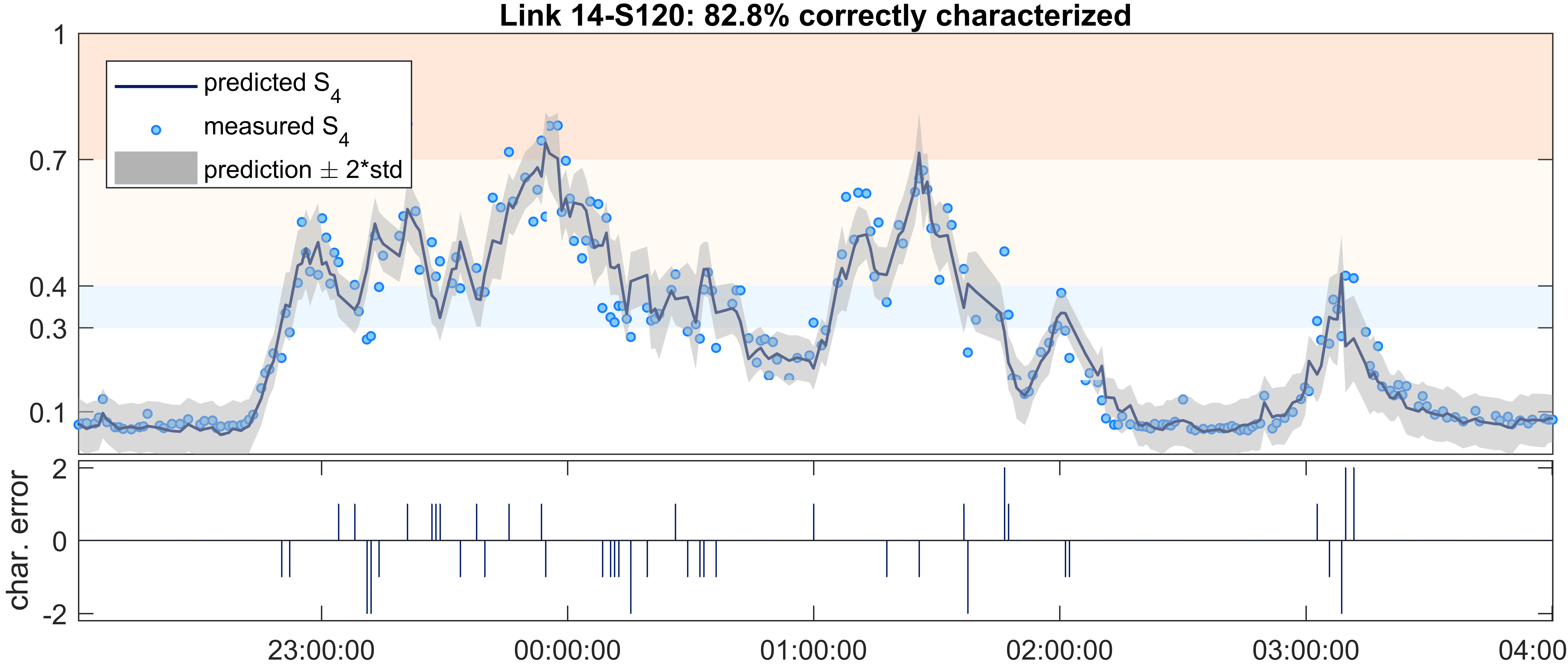}
           % \\ (a) Link: Receiver 14 and satellite S120 (SBAS)
            \\
            \includegraphics[width=0.4\columnwidth]{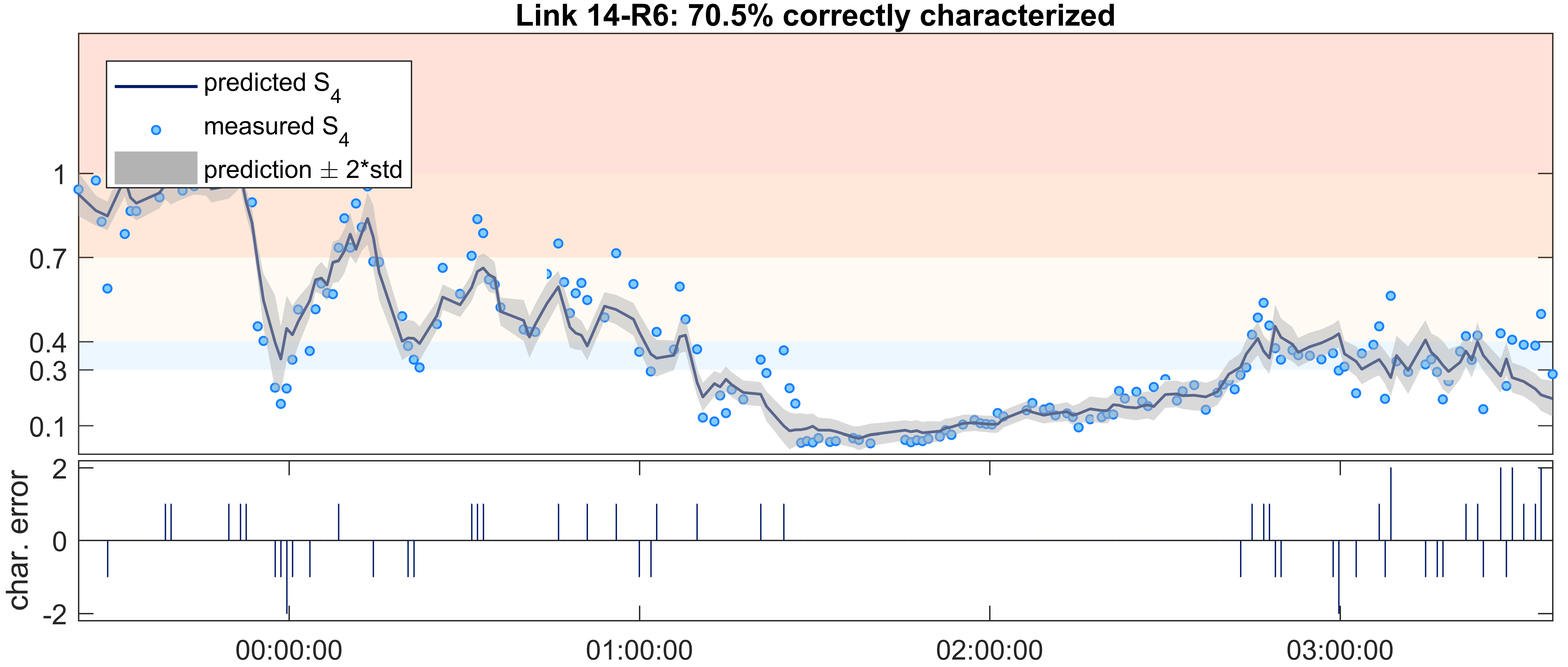}
            %\\ (b) Link: Receiver 14 and satellite R5 (GLONASS)
            \end{tabular}
%  \vspace{-1.2em}
 \caption{Comparisons between measurements and predictions for links between a ground receiver 14 and 2 satellites over the night Dec. 1-2, 2014.
 For each link, the top figure show the time series of the predicted $S_4$ values (black line) versus the measured values (blue circles). The grey area shows the two standard deviation interval around the predictions. The shaded colors depict the different scintillation characterisation levels (strong: $S_4>0.7$, moderate: $0.4<S_4\leq0.7$, weak: $0.3<S_4\leq0.4$, very weak: $S_4\leq0.3$). Under each time series is the discrepancy between predictions and measurements in terms of  scintillation characterization levels 1, 2 etc.}%
 \label{figure:ground14_sat}
\end{figure}

\begin{figure}[h]
 \centering
 \begin{tabular}{c}
         %   \includegraphics[width=1\columnwidth]{figures/link14s120_20210411.png}
           % \\ (a) Link: Receiver 14 and satellite S120 (SBAS)
        %    \\
        %    \includegraphics[width=1\columnwidth]{figures/link14r6_20210411.png}
            %\\ (b) Link: Receiver 14 and satellite R5 (GLONASS)
          %  \\
            \includegraphics[width=0.4\columnwidth]{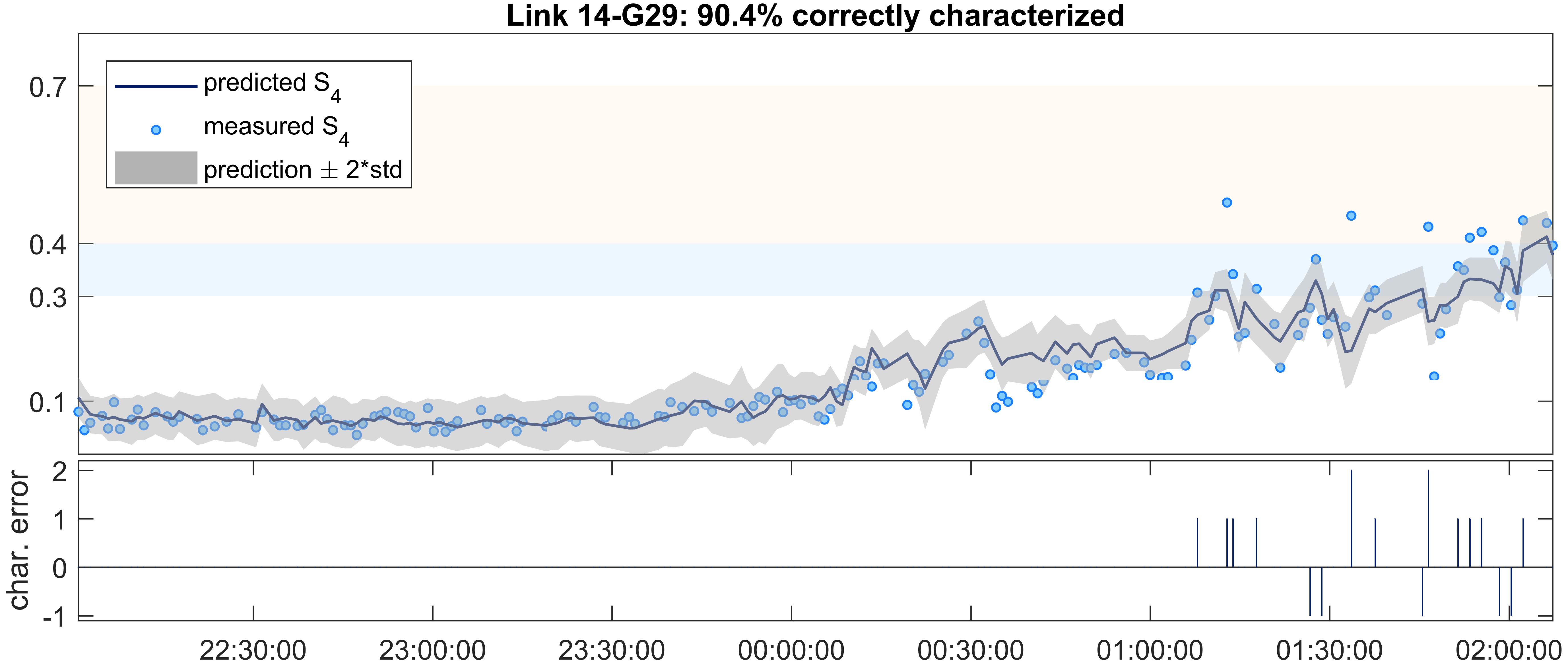}
           % \\ (c) Link: Receiver 14 and satellite G29 (GPS)
            \\
            \includegraphics[width=0.4\columnwidth]{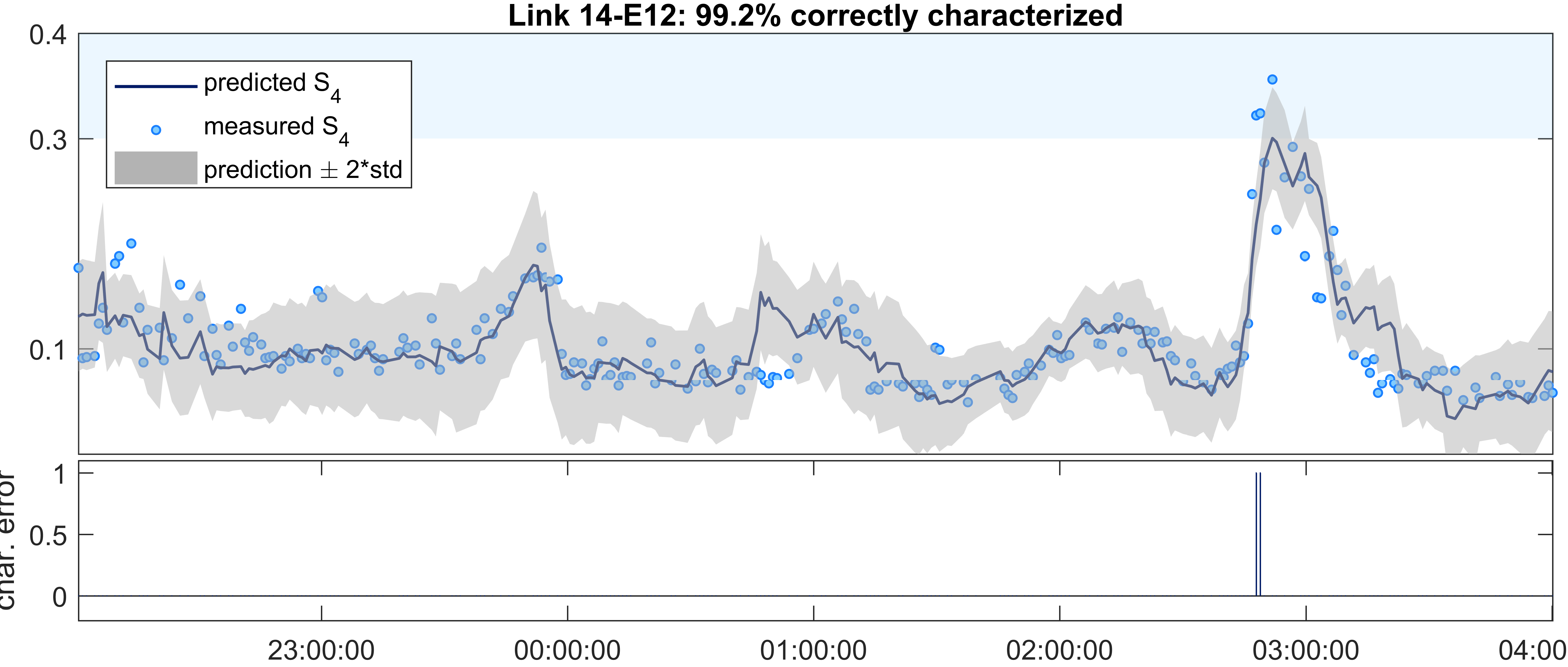}
           % \\ (d) Link: Receiver 14 and satellite E12 (Galileo)
 \end{tabular}
 \caption{Similar description as in Fig.~\ref{figure:ground14_sat} for the links R14-G29 (GPS) and R14-E12 (Galileo)}.%
 \label{figure:ground14_sat_2}
\end{figure}

In Figures ~\ref{figure:ground14_sat}  we present the results for links between receiver 14 and satellites S120 or R5 and in Figure \ref{figure:ground14_sat_2}, we have the links between receiver 14 and satellites G29 or E12 respectively. 
%Here, we present results for the link 14-S120  that were not used to produce the instant maps. 
Here,  S120 stands for satellite no 120 of the SBAS constellation, R5 for a satellite of GLONASS system, G29 for GPS system and E12 for Galileo system respectively.
  \vspace{-1.2em}
First, in the upper graph of Figure~\ref{figure:ground14_sat} we observe the accuracy of our images to predict the $S_4$ value of link 14-S120 by comparing the estimates (predictions) to the measured $S_4$ data (marked with light blue circles). Also, we plot the estimated 2 standard deviation interval around our predictions (in gray colour). This graph has been divided  horizontal zones with different colours which indicate the severity of scintillation activity. In particular, the white zone indicates the level where there is very weak scintillation activity ($0\leq S_4< 0.3$), the light blue zone designates the level with weak scintillation ($0.3\leq S_4<0.4$), the light yellow zone designates the level with moderate activity ($0.4<S_4\leq 0.7$), the light orange corresponds to the level with strong activity ($0.7\leq S_4\leq 1$) and the red zone is for values above 1 (the characterization zones are based on \cite{Veettil2020}).  Moreover, under the time series graph, we have an error bar graph that depicts the errors in scintillation characterization between the measurements and predictions at each time step $t$. In particular, when the observed measurements (validation data) and the prediction at time $t$ belong to the same scintillation zone then we say that the scintillation characterization (for example, strong, moderate, weak and very weak) is the same both for the prediction and the true measurement and then the characterization error is 0. On the other hand, when the characterization differs between the two data points then this error is quantified as level 1 or 2 (if a prediction gives lower scintillation activity than the measured data) and -1 or -2 (if the prediction is in a higher zone than the measured data). For instance, if a  prediction at time $t$ is located in the moderate zone and the measured data is located in the weak zone then the error is
-1. Moreover, above each time series graph we have included the percentage of the predictions that were in the correct scintillation zone.
to moderate scintillation activity our predictions are optimal (e.g. link 14-G29 or 14-E12) while there are some measurements ...
 Similar graphs we have for the links 14-G29 and 14-E12 in Figure~\ref{figure:ground14_sat_2}. 
Overall, we can observe that the predictions are in accordance with the actual measurements with small deviations and inaccuracies when abrupt changes  in the scintillation activity take place (e.g. large jumps in consecutive measurements) for all the tested links related to receiver 14. %More precisely, we see that for weak 
\FloatBarrier
\subsection{Receiver 53}
Similarly as in the previous subsection, we compare predictions with measurements for two links that have far less available data and scintillation monitors around them based on Figure~\ref{figure:Locations} and \ref{figure_used_data_locations} (right images). O
\begin{figure}[h]
\centering
\begin{tabular}{c}
           \includegraphics[width=0.4\columnwidth]{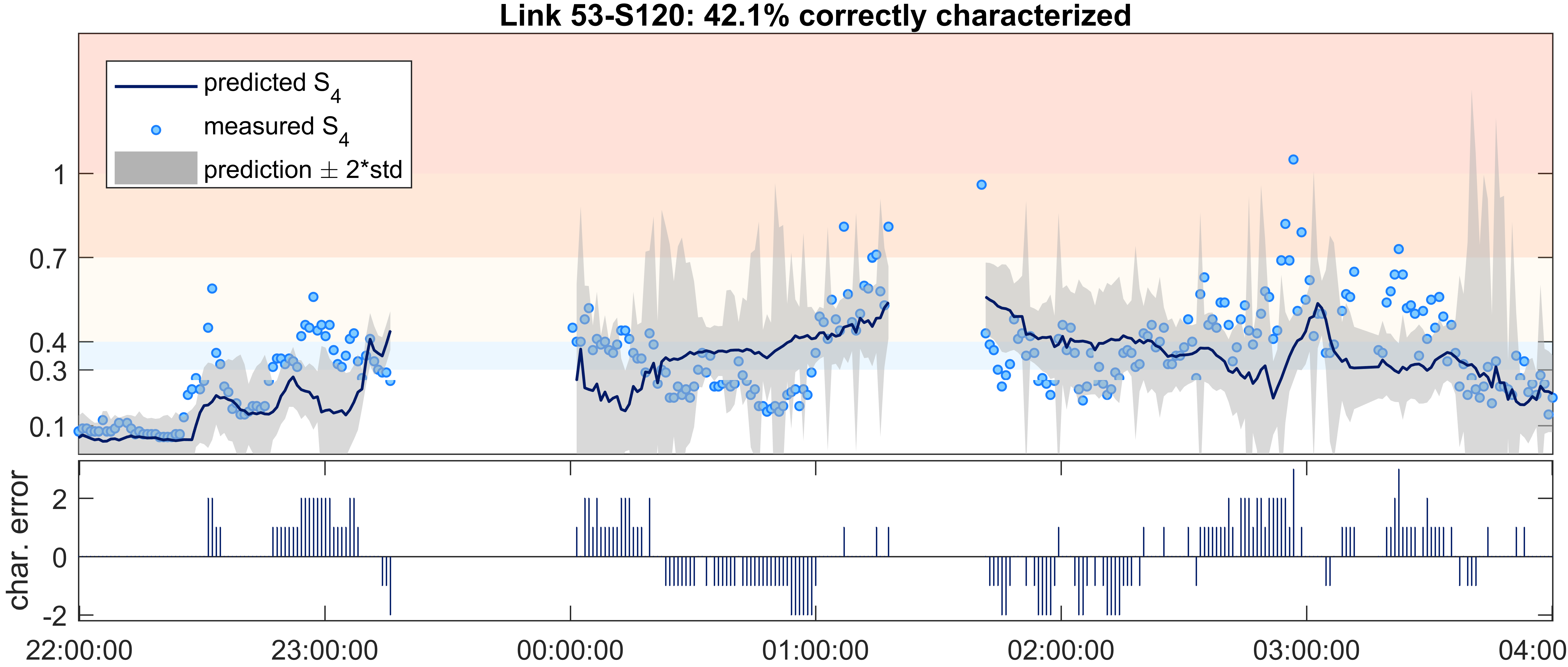}
          % \\ (a) Link: Receiver 53 and satellite S120 (SBAS)
           \\
          \includegraphics[width=0.4\columnwidth]{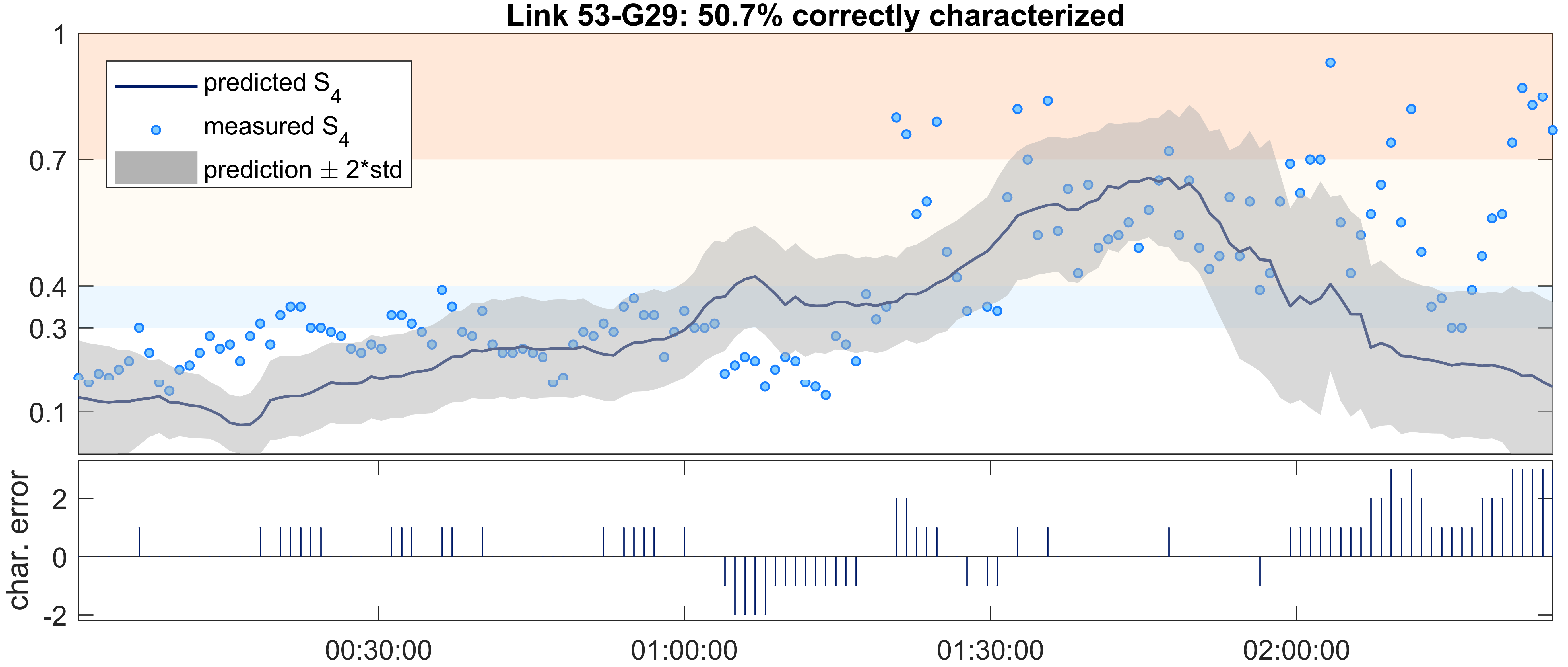}
        %  \\ (b) Link: Receiver 53 and satellite G29 (GPS)
\end{tabular}
 
\caption{Top and bottom figures are similar as in Fig. \ref{figure:ground14_sat}. Time series of the predicted $S_4$ values (black line) versus measured values (blue circles) for ground receiver 53 and 2 satellites (S120-SBAS and G29-GPS) during the night Dec. 1-2, 2014. }
\label{figure:ground53_sat}
\end{figure}
verall in Figure~\ref{figure:ground53_sat} we can observe that our predictions are less accurate when there is not enough available $S_4$ data in the vicinity of these links. For weak to moderate scintillation activity our algorithm works within a good tolerance. However, we can see large standard deviations and low percentages for correct scintillation characterization as it was expected due to the limited available data and difficulty to capture abrupt changes or high scintillation activity.  For the link 53-G29, we have that the prediction curve is relatively smooth following the measurements until 1.45 pm. However, after 2.00 am it misses the scintillation activity due to the lack of available coverage in this ionospheric region. Interestingly for the link 53-S120 (SBAS) we can observe that there are periods where there are not available measurements but unfortunately we are not aware why this is happening.
\FloatBarrier

\subsection{Validation metrics} 
In table~\ref{tab:metrics}, we quantify the results presented in Figures~\ref{figure:ground14_sat},~\ref{figure:ground14_sat_2} and \ref{figure:ground53_sat} with the help of the CM (\ref{eq:CM}) and RMS (\ref{eq:RMS}) metrics. As it was expected the correlation is large and the RMS is low for the links related to receiver 14 while the values of these metrics are large for receiver 53. Based on our results, we can conclude that the proposed approach can  produce images with good spatial and temporal resolution with high accuracy in these ionospheric areas where there is a good spatial and temporal coverage most of the time (i.e. small data gaps over time and relatively small spatial distance e.g. 1 or 2 units of  degree in longitude/latitude). This can be achieved by installing extra ground receivers and ideally by creating a dense spatially uniform distributed network of scintillation monitors which unfortunately currently is not available. This would be important for creating robust real time scintillation images everywhere over the continent which could provide critical information during the time interval where  strong scintillation activity takes place, especially at the anomaly.
\begin{table}[h]
\centering
\begin{tabular}{ |p{1cm}||p{1.3cm}|p{1.1cm}|p{1.1cm}| p{1.1cm}|p{1.3cm}|p{1.1cm}|  }
% \hline
 %\multicolumn{5}{|c|}{Country List} \\
 \hline
 Link & 14-S120 &  14-R5 & 14-G29 & 14-E12 &53-S120 & 53-G29\\
 \hline
 \hline
 CM &  0.95   & 0.94 &0.92 & 0.81& 0.39 & 0.44\\
 \hline
RMS&  0.062   & 0.097 & 0.043&0.03 &  0.18 &0.21\\ \hline
Cor.\% & 82.8  & 70.5  & 90.4 & 99.2 & 42.1 &50.7\\
  \hline
\end{tabular}
 % \vspace{-0.5em}
\caption{CM, RMS between the measurements and predictions and 
percentage of correct characterization in the predefined scintillation zones during the night time between 1-2 Dec. 2014 for the links presented in Fig.~\ref{figure:ground14_sat}, ~\ref{figure:ground14_sat_2} and ~\ref{figure:ground53_sat}. }\label{tab:metrics}
\end{table}
%\vspace{-1.5em}
\FloatBarrier

\subsection{Ensemble vs. single Kalman filter}
\begin{figure}[h]
\centering
\begin{tabular}{c}
           \includegraphics[width=0.4\columnwidth]{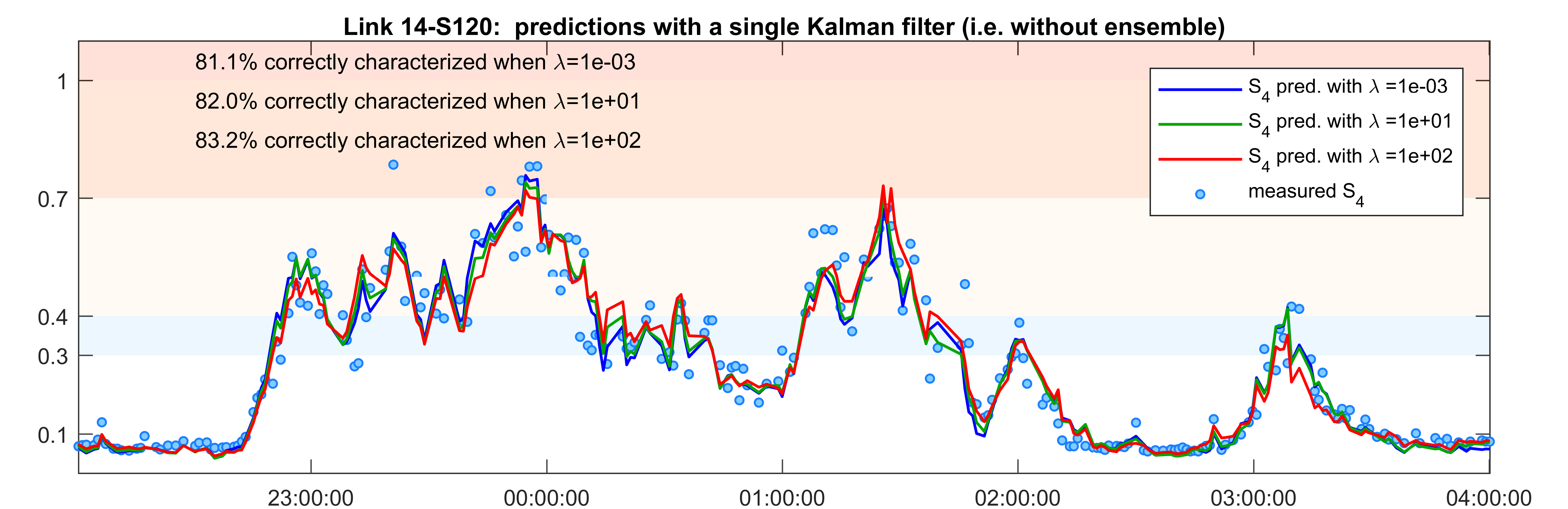}
         %  \\ (a) Link: Receiver 14 and satellite S120
           \\
           \includegraphics[width=0.4\columnwidth]{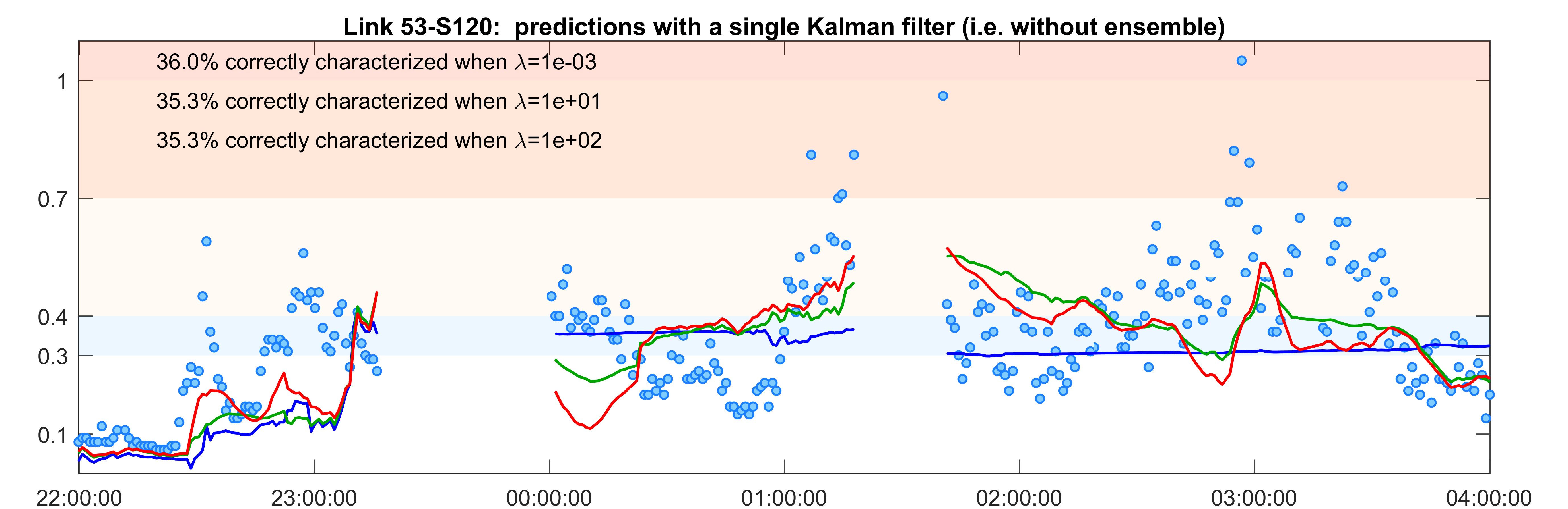} 
        %   \\ (b) Link: Receiver 53 and satellite S120
\end{tabular}

\caption{Predictions obtained by running a Kalman with a single tuning parameter $\lambda$.}
\label{figure:independant_l_53}
\end{figure}

Finally, we present the predictions that we would have obtained if we were running algorithm~\ref{alg:KalmanScintillation} with  single tuning parameters. Based on Figure~\ref{figure:independant_l_53}, we see that the prediction curves do not differ significantly when using single tuning parameters (instead of the ensemble) for the link 14-S120 due to the constant availability of observations. However, by comparing the percentage of correctly characterized scintillation activity for the link 53-S120, we have  that the ensemble (Figure~\ref{figure:ground53_sat}) performs better than the single kalman filters (Figure~\ref{figure:independant_l_53}).
 Moreover, for the link 53-S120 in Figure~\ref{figure:independant_l_53}, we can see a large variation in the predictions for single tuning parameters. This is because the Laplace smoothness (\ref{eq:laplace}) affects strongly the $S_4$ distribution at the ionospheric area around 53-S120. % in the absence of data. 
Large $\lambda$ will impose spatial smoothness on this area while small values force a constant value on the same area. %Finally, for the case of link: 14-S120, we see that the ensemble and the single tuning values result in similar predictions since there is constant availability of observations. 
% Finally, we can notice that the overall error percentages in scintillation characterization is a bit lower for the link....in Fi when employing in link....(Figure~\ref{figure:ground53_sat})the ensemble compared  to single regulation values $\lambda$ (as appear in Figure~\ref{figure:independant_l_53}).
%\FloatBarrier
%%%%%%%%%%%%%%%%%%%%%%%%%%%%%%%%%%%%%%%%%%%%%%%%%%%%%%%%%%%%%%%%%%%

\section{Conclusions and future work}\label{sec:conclusions}

In this work, we proposed an algorithm that employed the Kalman modelling in conjunction with numerical and learning techniques to produce ionospheric $S_4$ images of high temporal and spatial resolution. We demonstrated our approach by producing scintillation images over South America  during the night-time 1-2 December 2014 which was a period of mild to strong scintillation activity. We validated the accuracy of our images by comparing estimates (based on our images) with measurements that were not used to produce these images. Our results saw that we can predict with a very high accuracy the $S_4$ values (i.e. both weak and strong activity) in  ionospheric areas where there is a fairly good coverage by ground receivers. We remark that our approach can be directly applied for scintillation monitoring (e.g. amplitude, phase or variance of TEC) over other areas which are suffering of communication disruptions due  to scintillation e.g. over the Scandinavian peninsula and Finland. Furthermore, to increase the accuracy of our images, our approach could include other data e.g. occultation  measurements \cite{Groves2009,Wu2020}.
Regarding our proposed framework, there are several modifications which could be considered in the future. For example, we could 
update dynamically the set of the tuning parameters as in \cite{Chaer1997}. Moreover, the noise covariance matrices in (\ref{eq:evolutionModel_walk}) and  (\ref{eq:ObservationModelFinal})
could be estimated using  learning techniques \cite{Murphy2012}. Another point is the production of 3D images or to introduce another matrix $L$ instead of (\ref{eq:NormLaplace}); however
all the previous ideas require extra data which  could be acquired if more scintillation receivers and satellites become available in the future.

%\input{MainText}

%\appendix

%\input{Appendix_shrink}

\section*{Acknowledgement}
The work was supported by the Academy of Finland Postdoctoral Researcher
program (No 316542). I would like to cordially thank Dr. N. D. Smith for his suggestions how to handle the unknown tuning parameter, providing comments and our fruitful discussions. Also, I would like to thank Dr. V. Rimpil\"ainen for reading my manuscript, proposing improvements and helping to edit the results sections.

%\FloatBarrier

\bibliographystyle{plain}
%\bibliographystyle{bib/my_phd_bib_style}
%\bibliographystyle{IEEEtran}
%\bibliography{bib_scintillation_v1}
\bibliography{main_arxiv}
%\appendix
\section*{Appendices}
\subsection{Predictive distribution}\label{Appendix_1}
 In general, we denote random variables by capital
letters and their realizations by lowercase letters. Here,
$\{S_t\}_{t=0}^\infty$ and $\{Y_t\}_{t=0}^\infty$ are two Markov
stochastic processes where the random vector $S_t\in\mathbb{R}^N$ represents
the quantity of primary interest (i.e. ionospheric scintillation
coefficients) and is called the state vector, and
$Y_t\in\mathbb{R}^{M_t}$ is the observation vector at %time instant
$t$.
For a state space model, Markov
process properties are introduced. In particular, the properties are i)
$\{S_t\}_{t=0}^\infty$ is Markov process i.e. the conditional
probability density $\pi(s_t|s_{t-1},\ldots,s_0) =
\pi(s_t|s_{t-1})$, ii) $\{Y_t\}_{t=0}^\infty$ is Markov process with
respect to the history of $\{S_t\}_{t=0}^\infty$ which means that
 $\pi(yt|s_t,s_{t-1},\ldots,s_0) =
\pi(y_t|s_t)$, and iii) process $\{S_t\}_{t=0}^\infty$ depends on
the past observations only through its own history, i.e.
 $\pi(s_t|s_{t-1},y_{t-1},\ldots,y_1) =
\pi(s_t|s_{t-1})$ %. Please find further details in 
\cite{2005Kaipio}.

The predictive distribution  $\pi(s_t|D_{t-1})$ (\ref{eq:historyprior}) 
can be computed by the Chapman-Kolmogorov equation \cite{Murphy2012}
\begin{equation*}\label{eq:evolution-based predicted density}
\pi(s_t|D_{t-1}) = \int \pi(s_t|s_{t-1})\; \pi(s_{t-1}|D_{t-1})\;
ds_{t-1},
\end{equation*}
where $ \pi(s_t|s_{t-1})$ is the transition probability and depends
on the evolution model (\ref{eq:evolutionModel_walk_1}), and $\pi(s_{t-1}|D_{t-1})$ is the posterior
density of the previous time step. Under Gaussian condition for the noise and the prior,  the posterior
is Gaussian i.e. $\pi(s_{t-1}|D_{t-1}) \sim
\mathcal{N}(\hat{s}_{t-1},\Gamma_{s_{t-1}|D_{t-1}})$, where
$\hat{s}_{t-1}$ is the mean
 and $\Gamma_{s_{t-1}|D_{t-1}}$ is the
posterior covariance at time  $t-1$.

 \begin{equation*}
\pi(s_t|D_{t-1})  \propto \\
 \int  \exp{\left(-\frac{1}{2}
 (s_t-s_{t-1})^\mathrm{T}\Gamma_{n_t}^{-1}
 (s_t-s_{t-1}) 
 \right)}
 \exp{\left(-\frac{1}{2}
 (s_{t-1}-\hat{s}_{t-1})^\mathrm{T}\Gamma_{s_{t-1}|D_{t-1}}^{-1}
 (s_{t-1}-\hat{s}_{t-1}) \right)}\;ds_{t-1}.
 \end{equation*}

\subsection{Coefficient of the observation matrix}\label{Appendix_2}

To estimate the coefficients of matrix $A_t$, %w consider
%the problem in an element-wise manner. In particular
we first find the element where the IPP
$\mathrm{x}_j(t)=(x^{lat}_j,x_j^{lon})$ of observation
$y_t(\mathrm{x}_j)$ is projected to. If point $\mathrm{x}_j(t)$ lies
on element $\mathcal{T}_k$ with nodes $\mathrm{x}_{p_k}$
($k=1,2,3$)\footnote{This is a local indexing to distinguish the three
nodes of element $\mathcal{T}_k$}, then the observation
$y_t(\mathrm{x}_j)$ is expressed as a linear combination of the
three basis functions which have support on this triangular element
denoted by $\phi_{p_k}(\mathrm{x})=a_{p_k}
x^{lat}+b_{p_k}x^{lon}+c_{p_k}$. Then,
$y_t(\mathrm{x}_j)=\sum_{k=1}^3\phi_{p_k}(\mathrm{x}_j(t)) s_{p_k}$,
where $s_{p_k}$ are the scintillation values at the corresponding
element's nodes $\mathrm{x}_{p_k}$. Now, the coefficients $\{a_{p_k},b_{p_k},c_{p_k} \}$ of the basis
functions can be estimated by solving a set of linear equations where
$\phi_{p_k}(\mathrm{x}_{p_l})=1$ when $l=k$ and
$\phi_{p_k}(\mathrm{x}_{p_l})=0$ when $l\neq k$. Then we estimate
$\phi_{p_k}(\mathrm{x}_j(t))=a_{jk}$. Hence $S_4$ value
$y_t(\mathrm{x}_j)$ is a weighted sum of the $S_4$ values of the
nodes of the element where $\mathrm{x}_j$ is located.  %(based on the
%location $\mathrm{x}_j$ the three weights $a_{jk}$ are determined).
An alternative way
to estimate these coefficients is to use the isoparametric mapping
\cite{Hughes2000}.

\end{document}